\documentclass[twocolumn,doublespacing]{aastex631} 
\usepackage{amsmath}
\usepackage{mathtools}
\usepackage{appendix}
\usepackage{dsfont}
\usepackage{listings}
\usepackage{placeins}
\usepackage{booktabs}

\usepackage{hyperref}

\shorttitle{{Tidal Deformation of Minor Bodies }}
\shortauthors{Taylor et al.}

\graphicspath{{./}{figures/}}

\begin{document}

\title{Numerical Simulations of Tidal Deformation and Resulting Light Curves of Small Bodies: Material Constraints of 99942 Apophis and 1I/`Oumuamua}

\correspondingauthor{Aster Taylor}
\email{astertaylor@uchicago.edu}
\author[0000-0002-0140-4475]{Aster G. Taylor}
\affiliation{Dept. of Astronomy and Astrophysics, University of Chicago, 5640 S Ellis Ave, Chicago, IL 60637} 

\author[0000-0002-0726-6480]{Darryl Z. Seligman}
\affiliation{Dept. of Astronomy and Carl Sagan Institute, Cornell University, 122 Sciences Drive, Ithaca, NY, 14853, USA}

\author[0000-0003-0647-6176]{Douglas R. MacAyeal}
\affil{Dept. of Geophysical Sciences, University of Chicago, 5734 S Ellis Ave, Chicago, IL 60637}

\author[0000-0001-6952-9349]{Olivier R. Hainaut}
\affiliation{European Southern Observatory, Karl-Schwarzschild-Strasse 2, Garching bei München, D-85748, Germany}

\author[0000-0002-2058-5670]{Karen J. Meech}
\affil{Institute for Astronomy, University of Hawaii, 2680 Woodlawn Drive, Honolulu, HI 96822, USA}

\begin{abstract}

In this paper, we present an open source software ({$\texttt{SAMUS}$}) which simulates constant-density, constant-viscosity liquid bodies subject to tidal forces for a range of assumed viscosites and sizes. {This software solves the Navier-Stokes equations on a finite-element mesh, incorporating the centrifugal, Coriolis, self-gravitational, and tidal forces. The primary functionality is to simulate the deformation of minor bodies under the influence of tidal forces. It may therefore be used to constrain the composition and physical structure of bodies experiencing significant tidal forces, {such as 99942 Apophis and 1I/`Oumuamua. We demonstrate that $\texttt{SAMUS}$ will be useful to constrain the material properties of Apophis during its near-Earth flyby in 2029. Depending on the material properties, Apophis may experience an area change of up to 0.5\%, with similar effects on the photometric brightness. We also} apply $\texttt{SAMUS}$ to constrain the material dynamic viscosity of 1I/`Oumuamua, the first interstellar object discovered traversing the inner Solar System. `Oumuamua experienced a close approach to the Sun at perihelion ($q\simeq 0.25$ au) during which there were significant tidal forces that may have caused deformation of the body. This deformation could have lead to observable changes in the photometric light curve based on the material properties. The application of \texttt{SAMUS} to produce} synthetic observations which incorporate tidal deformation effects demonstrate that no deformation {--- an infinite dynamic viscosity ---} best reproduces the photometric data. {While these results indicate} that `Oumuamua did not experience significant tidal deformation, {a sophisticated model incorporating non-principal axis rotation is necessary to conclusively analyze both `Oumuamua and Apophis.}
\end{abstract}

\keywords{Interstellar Objects (52); Comets (280); Hydrodynamics (1963)}

\section{Introduction}

{Tidal gravitational forces are one of the primary drivers of the evolution of the  objects within the Solar System. For example, the geosynchronous orbit of the Moon was explained by \citet{Darwin1879,Darwin1880} as a natural consequence of tidal dissipation. The theory of spin  evolution of satellites due to tidal forces was improved vastly in the following centuries \citep{Kaula1964,MacDonald1964}. The discovery of the 3:2 spin orbit resonance of Mercury \citep{Pettengill1965} led to significant theoretical advances on long-term tidal dissipation \citep{Peale1965,Liu1965,Colombo1965,Goldreich1966:spinorbit,Goldreich1968,Bagheri2022}. }

{The theory of tides has led to multiple predictions that were later corroborated by data. An archetypal example of this was the prediction of vulcanism on Io by \citet{Peale1979} which was later verified by the Voyager 1 spacecraft \citep{Smith1979}. Similarly, \citet{Wisdom1984} predicted that the Saturnian satellite Hyperion was chaotically tumbling, which was confirmed  by Voyager 2  \citep{Black1995}. For a more recent example, \citet{Fuller2016} predicted the rapid outward migration of the Saturnian satellite Titan due to sustained excitation of inertial waves via tidal forces, which was then measured by the Cassini spacecraft \citep{Lainey2020}.} 

{Tidal deformation can also affect small bodies like comets and asteroids, and in the most dramatic cases, lead to catastrophic disruption events. A historical example of this was the tidal disruption of the comet Shoemaker-Levy 9 into Jupiter's atmosphere \citep{Shoemaker1993, Weaver1995,Lellouch1995,Noll1995}. Chains of craters on the surfaces of Callisto and Ganymede have also been explained by similarly catastrophic tidal events \citep{Schenk1996}. In milder cases tidal deformation can result in satellite formation via material stripping \citep{Richardson1998}, which has been invoked to explain the near ubiquity of doublet craters on Solar System bodies \citep{Melosh1991,Bottke1996,Bottke1996b,Melosh1996,Cook2003}. In this paper, we consider the mildest case, where tidal forces lead to body deformation, with application to the near-Earth object 99942 Apophis and the interstellar object 1I/`Oumuamua \citep{mpec2017,Williams17}. For recent reviews on interstellar objects, see \citet{Jewitt2022ARAA} and \citet{MoroMartin2022}. }

{Apophis was discovered in 2004 and subsequently identified as a potential Earth impact threat. While follow-up radar observations have eliminated any chance of impact within a century, Apophis will pass within $\sim$6 Earth radii in 2029 \citep{Brozovic2018}, providing an excellent opportunity for observation and characterization. The OSIRIS-REx mission, after returning from the asteroid 101955 Bennu, will be renamed OSIRIS-APEX and dispatched to encounter and observe Apophis \citep{Nolan2021}. Radar observations \citep{Brozovic2018} revealed that Apophis has a bi-lobed shape. Observations also revealed that Apophis exhibits non-principal-axis rotation  \citep{Brozovic2018,Pravec2014,Lee2022}, and measured a non-zero Yarkovsky effect \citep{Perez2022}.}

{The potential effects of Apophis' near-Earth flyby on the geophysical, photometric, and rotational states have been investigated. \citet{Benson2023} demonstrated that the close encounter will induce sufficient tidal torques to modify the rotational state. The tidal forces will also be sufficient to produce resurfacing events which could modify the photometric properties \citep{Yu2014,Kim2023}. The magnitude and nature of these effects are potentially effective probes of the geophysical properties of Apophis, including density distribution \citep{Dinsmore2022}, rigidity and dissipation \citep{Hirabayashi2022}, and seismic response \citep{DeMartini2019}. We investigate the material properties for which tidal forces  may  induce a detectable shape deformation in the object, enabling \texttt{SAMUS} to constrain the material properties of Apophis. Similar effects are investigated for I1/`Oumuamua.}

{After the October 2017 discovery of `Oumuamua, there} was an immediate acquisition of ground- and space-based observations of the rapidly fading object. These observations produced a high-quality composite light curve spanning approximately a month (29.3 days) and a spatial segment of $l\simeq0.13$ au. In total there were 818 observations, reported by \citet{meech2017}, \citet{bolin2017}, \citet{bannister2017}, \citet{drahus2017}, \citet{fraser2018}, \citet{jewitt2017}, \citet{knight2017}, and \citet{belton2018}. These observations were {collectively reported} in \citet{belton2018}. 

`Oumuamua was interpreted to be highly elongated, with an aspect ratio estimated to be $>$3:1 \citep{bolin2017,knight2017}, $>$5:1 \citep{bannister2017,fraser2018,jewitt2017}, and up to 10:1 \citep{meech2017}. Frequency analysis of the light curve showed a maximum at a period of $p\simeq4.3$ hours \citep{belton2018}. This was interpreted to be half of the rotational period --- corresponding to a revolution of 180$^\circ$. `Oumuamua's variations in absolute magnitude of $H\simeq22.5\pm1.3$ over its rotation led to the conclusion that it was exhibiting complex, non-principal axis rotation \citep{drahus2017,meech2017,fraser2018}.

This analysis was further refined by \citet{mashchenko2019} who demonstrated via full light curve modeling that a near-symmetric oblate ellipsoid with dimensions of 115:111:19$\sim$6:6:1 meters provided a best fit geometry for the light curve data. This size estimate assumes a geometric albedo of $A=0.1$ and would change with a different albedo, but with appropriately scaled dimensions. While a prolate ellipsoid with dimensions of 342:42:42 meters is also allowable, the torques required to replicate the motion are highly tuned, so the prolate geometry is disfavored. 

Deep imaging revealed a notable lack of cometary activity, classifying `Oumuamua as an asteroidal body and restricting possible dust outputs (upper limits range from $\sim2\cdot10^{-4}$ kg s$^{-1}$ \citep{jewitt2017} to $1.7\cdot10^{-3}$ kg s$^{-1}$ \citep{meech2017}). Additionally, while outbound at 2 au, there was a significant non-detection of the object with the \textit{Spitzer Space Telescope}. This non-detection placed limits on the production of CO and CO$_2$ \citep{trilling2018}. 

Astrometric positional data revealed that the trajectory was inconsistent with pure Keplerian motion \citep{micheli2018}. The addition of a radially outward non-gravitational acceleration of the form $a=4.92\cdot10^{-4}\, (r/1\, {\rm au})^{-2}\, \boldsymbol{\hat{r}}$ cm s$^{-2}$ provides a greatly improved, $30\sigma$ fit to the trajectory.\footnote{An acceleration with a form of $r^{-1}$ is nearly as good of a fit.} \citet{micheli2018} proposed a comet-like outgassing as an explanation for this acceleration, ruling out radiation pressure, the Yarkovsky effect, magnetic forces, and others which would require extreme physical properties.

The restrictions on the coma and micron-scale dust presence in the vicinity are quite stringent. Therefore, theories for the provenance of the object positing cometary outgassing as the source of the acceleration require additional complexity to avoid violating the \textit{Spitzer} or photometric observations.\footnote{Although the original \textit{Spitzer} estimates had a computational error, see \citet{seligman2021}. Even these revised CO limits, however, are prohibitive to the $1/r^2$ fit.} H$_2$O ice was initially proposed as an outgassing accelerant because it is the most common volatile in Solar System comets \citep{Rickman2010,Ahearn2012,Ootsubo2012,Cochran2015,Biver2016,Bockelee2017} and its presence is not in tension with the \textit{Spitzer} non-detection. However, the relatively high enthalpy of sublimation (51 kJ mol$^{-1}$) of H$_2$O implies that water sublimation would require more energy input than `Oumuamua received from solar radiation \citep{sekanina2019}.

Attempting to unify these constraints, \citet{SL2020} argued that only hypervolatiles could serve as the accelerant for `Oumuamua. They found that only molecular hydrogen (H$_2$), neon, molecular nitrogen (N$_2$), and argon were allowable accelerants for an oblate spheroid --- although CO was also shown to be energetically feasible. Those authors also investigated the feasibility of hydrogen ice as the bulk constituent --- originally hypothesized by \citet{fuglistaler2018solid} --- as it requires the lowest active surface fraction to be explanatory. In this hypothesis, `Oumuamua would have formed in a failed prestellar core in a Giant Molecular Cloud. This model naturally explains the extreme shape (via continuous H$_2$ ablation), the low excess velocity speed, and young age \citep{mamajek2017,Gaidos2017a, Feng2018,Fernandes2018,hallatt2020,Hsieh2021}. However, there are theoretical barriers to the formation of macroscopic bodies composed of solid hydrogen, such as the frigid temperatures required for formation and rapid evaporation in the interstellar medium \citep{hoang2020,phan2021,levine2021,LL2021}. Although the low condensation temperature of molecular hydrogen ($<$10K) poses difficulties for its formation, \citet{LL2021} demonstrated that adiabatic expansion pockets were a plausible formation environment for such an object. \citet{jackson2021} instead suggested that `Oumuamua was composed of molecular nitrogen (N$_2$) ice, while \citet{desch2021} proposed that impacts on extrasolar Pluto analogues would provide a plausible source for objects like `Oumuamua. However, \citet{levine2021} demonstrated that the necessary mass density for this formation to be plausible is unreasonably high. \citet{seligman2021} found that a typographical mistake in \citet{trilling2018} led to the reported outgassing limits of CO to be underestimated by two orders of magnitude. When this error was corrected for, those authors showed that a body characterized by a modest covering fraction of CO exhibiting sporadic activity could explain both the acceleration and Spitzer observations. 

\citet{Bergner2023} demonstrated that the crystallization of amorphous water ice would produce sufficient radiolytically produced and entrapped H$_2$ to provide the observed nongravitational acceleration. This crystallization would occur in the absence of sublimation of the overall ice matrix, thereby explaining the lack of {dust coma observed. Further more, \citet{Chesley2016}, \citet{Farnocchia2022} and \citet{Seligman2022} reported statistically significant non-gravitational accelerations on seven near-Earth objects (NEOs) that did not display visible activity, similar to `Oumuamua. 
 
{Motivated by these advances, we reconsider the hypothesis} that `Oumuamua {was an icy object. \citet{flekkoy2019} considered the effects of tidal forces in the context of a dust aggregate and demonstrated a surprising stability against tidal stresses. We build upon this work to consider} shearing effects due to tidal forces in the context of an icy cometary body. At perihelion, `Oumuamua passed within $r_H\simeq0.256$ au of the Sun, subjecting it to non-trivial tidal stresses --- although attempts to pre-discover the object here resulted in non-detections \citep{hui2019}. The tidal stresses, combined with the large-magnitude centrifugal force produced by `Oumuamua's high aspect ratio and rapid rotation, produce significant shearing stress on the body. In this paper, we investigate the effects of these shearing forces and provide constraints on the size and dynamic viscosity of `Oumuamua. The latter could potentially provide a constraint on the possible material composition of `Oumuamua.

\section{{Numerical Simulations of Tidal Deformation}}\label{sec:numerics}

Many small asteroids are loosely-bound granular bodies. Characterization of plastic deformation of these bodies under rotational \citep{hirabayashi2014,hirabayashi2015,hirabayashi2019} and tidal stress \citep{kim2021} has been the subject of much research, and numerical models similar to those developed and presented in this paper are often applied to these objects \citep{hirabayashi2019,sanchez2017}. Numerical calculations incorporating finite-element and soft-sphere discrete approximations are routinely used to analyze the tensile strength of these objects with and without cohesive forces. In this paper, we instead model {these objects as fluid masses,} which greatly simplifies the analysis. While this approximation is a simplification, it is appropriate for both semi-crystalline solids such as ices and --- in certain circumstances --- for granular materials subjected to cohesive and friction forces. This simplification reduces the problem to only two degrees of freedom --- the dynamic viscosity $\mu$ and the density $\rho$ --- which fully describe the material properties and state of `Oumuamua. This simplification enables us to model the deformation explicitly over time, instead of relying on the `deformation modes' identified in \citet{hirabayashi2019}. 

{In this section, we present a generalized software, Simulator of Asteroid Malformation Under Stress (\texttt{SAMUS}), which simulates the} deformation of {constant-density constant-viscosity liquid-body ellipsoids under forcing pressures, which can be applied to constrain the dynamic viscosity and size of minor objects. \texttt{SAMUS} incorporates tidal, centrifugal, Coriolis, and self-gravitational forces for minor bodies, and allows for customized trajectory, principal axes, rotational period, density, dynamic viscosity, rotational axis, and simulation cutoffs. It is accessible on \href{https://pypi.org/project/SAMUS/1.0.0/}{PyPi} and at \citet{SAMUS}, and can be installed via \texttt{pip}.}

{\texttt{SAMUS} currently implements several simplifying assumptions due to computational practicalities. \texttt{SAMUS} uses a {``}fixed-axis{"} NPA rotation and a constant rotational period; tumbling and 3-dimensional rotation are not currently included. However, this is not stringently enforced in the model. The rotation axis is used to compute the non-inertial forces and to calculate the tidal force in the body frame. However, \texttt{SAMUS} does allow the orientation of the body to be shifted by the tidal forces, allowing for a more complex and accurate evolution. However, it is well-known that tidal forces can induce angular momentum axis drift, and so the incorporation of that effect into \texttt{SAMUS} will be incorporated in future versions of this software --- currently, the rotational axis and magnitude are kept fixed, and the angular momentum evolves with the changing shape.}

{\texttt{SAMUS} also assumes that the simulated object is not subjected to ablation, and that shape changes only arise from specified shearing forces. This package also assumes that the simulated body is homogeneous, both in density and in dynamic viscosity. Given the temperature and material dependence of both of these properties, this implies a constant temperature and material over the body --- at least across the scale of the cell size. Crystallization of amorphous materials would induce (minor) changes in the viscosity of the material and change the temperature, but this effect is currently not included in these simulations. }

\subsection{{Numerical Calculations}}
 
{The \texttt{SAMUS} simulation software is written in Python 3.8.10 \citep{python3}, and is primarily based on the \texttt{FEniCS} \citep{fenics1,fenics2}, \texttt{UFL} \citep{UFL}, and \texttt{DOLFIN} \citep{dolfin1,dolfin2} packages. It also has dependencies on \texttt{NumPy} \citep{numpy}, \texttt{SciPy} \citep{scipy}, \texttt{pandas} \citep{pandas1,pandas2}, \texttt{quaternion} \citep{quaternion}, and \textit{MPI for Python} \citep{mpi1,mpi2,mpi3,mpi4}. All of these must be installed in the user's distribution. \texttt{SAMUS} is primarily structured as a Python class, and solves the weak formulation of the partial differential Navier-Stokes equations over a finite-element mesh.}

{The domain used by \texttt{SAMUS} is an $\mathcal{S}_3$ (3-ball) domain (created by Gmsh \citep{gmsh} and loaded into \texttt{DOLFIN}), distorted into an ellipsoid with principal axes $a,b,c$. After reading in the provided trajectory data, body parameters, and simulation parameters, \texttt{SAMUS} uses an Euler finite-difference approximation to iteratively solve the Navier-Stokes equations. The mesh is advectively updated at each time step to simulate the tidal deformation, where the computed fluid velocity is used to find the displacement vector. }

{\texttt{FEniCS} is used to solve the weak formulation of the incompressible Navier-Stokes equations with Dirichlet boundary conditions. In the following equations, $\boldsymbol{u}=\partial\boldsymbol{r}/\partial t$ is the velocity, $\boldsymbol{r}$ is the position in the co-moving, non-inertial frame, $\rho$ is the density, $\mu$ is the dynamic viscosity, $p$ is the pressure, and $\boldsymbol{\Omega}$ is the angular velocity vector. $V$ and $Q$ are function spaces over $\mathds{R}^3$ and $\mathds{R}$ respectively,\footnote{Defined as continuous Galerkin domains.} with $\boldsymbol{u},\boldsymbol{v}\in V$ and $p,q\in Q$, and $\boldsymbol{v},q$ are test functions. The differential `$\text{d}x$' represents a volume integral over the body domain. }

{The strong momentum equation is 
\begin{equation}
    \rho\frac{\partial\boldsymbol{u}}{\partial t}+\rho(\boldsymbol{u}\cdot\nabla)\boldsymbol{u}-\nabla\cdot\boldsymbol{\sigma}(\boldsymbol{u},p)=\boldsymbol{f}\,.
\end{equation}
\texttt{SAMUS} uses the weak form, which is}
\begin{equation}\label{eq:momentum}
\begin{aligned}
    &\rho\frac{\partial\boldsymbol{u}}{\partial t}\cdot\boldsymbol{v}\ \text{d}x + \mu \nabla\boldsymbol{u}\cdot\nabla\boldsymbol{v}\ \text{d}x\\
    +&\rho(\boldsymbol{u}\cdot\nabla)\boldsymbol{u}\cdot\boldsymbol{v}\ \text{d}x-p\nabla\cdot\boldsymbol{v}\ \text{d}x\\
    =&\boldsymbol{F}_{\text{tidal}}\cdot\boldsymbol{v}\ \text{d}x+\boldsymbol{g}\cdot\boldsymbol{v}\ \text{d}x\\
    -&\rho (\boldsymbol{\Omega}\times(\boldsymbol{\Omega}\times\boldsymbol{r}))\cdot\boldsymbol{v}\ \text{d}x\\
    -&2\rho(\boldsymbol{\Omega}\times\boldsymbol{u})\cdot\boldsymbol{v}\ \text{d}x\ \ \ \forall\boldsymbol{v}\in V.
\end{aligned}
\end{equation}
The {mass continuity equation, in its strong form, is
\begin{equation}
    \nabla\cdot\boldsymbol{u}=0,
\end{equation}
and \texttt{SAMUS} again uses the weak form,
\begin{equation}\label{eq:masscont}
    q\nabla\cdot\boldsymbol{u}\ \text{d}x=0\ \ \ \forall q\in Q.
\end{equation}}

{The derivation of the weak form from the strong form of the Navier-Stokes equations is given in \citet{quarteroni2014}. The right-hand side of the momentum equation represents the forcing, each term of which is defined in Table \ref{table:forcing}.}
 
\begin{table}[ht]
\begin{tabular}{ |l|c| } 
 \hline
 \multicolumn{2}{||c||}{FORCING TERMS} \\
 \hline\hline
 Tidal force:&$\boldsymbol{F}_{\text{tidal}} $\\ 
 \hline
 Self-gravitational force:&$\boldsymbol{g} $\\
 \hline
 Centrifugal force:&$-\rho(\boldsymbol{\Omega}\times(\boldsymbol{\Omega}\times\boldsymbol{r}))$\\
 \hline
 Coriolis force:&$-2\rho(\boldsymbol{\Omega}\times\boldsymbol{u})$.\\
 \hline
\end{tabular}

\caption{The forcing terms in the Navier-Stokes equations.}

\label{table:forcing}
\end{table}
 
{In \texttt{SAMUS}, the acceleration $\partial\boldsymbol{u}/\partial t$ is estimated with an Euler finite-difference method. For a given time step indexed by $i$, the acceleration is approximated as $\partial\boldsymbol{u}/\partial t\simeq(\boldsymbol{u}_i-\boldsymbol{u}_{i-1})/\Delta t$. The time step $\Delta t$ is adaptively modified to ensure that the Courant–Friedrichs–Lewy (CFL) condition ($|\boldsymbol{u}|\Delta t/\Delta x<C_{\text{max}}$) is met \citep{CFL}. $C_{\text{max}}$ can be user-defined and is set to a default of 1, which is the standard limit. We performed extensive stability and convergence tests, which are available in the \texttt{SAMUS} package (and not described in this paper).}

{\texttt{SAMUS} additionally uses \texttt{FEniCS} to rapidly solve the weak form of the self-gravitational force as given by Gauss, 
\begin{equation}\label{eq:gaussgrav}
    c\nabla\cdot\boldsymbol{g}\ \text{d}x=-4\pi G\rho\ c\ \text{d}x\ \ \ \forall c\in Q\,,
\end{equation}
where $c$ is a scalar test function in $Q$. This method is relatively rapid and allows for efficient computation of self-gravity for even highly distorted bodies.}
 
{\texttt{SAMUS} produces a }\verb|csv| {file containing timestamps, the maximum dimension of the body on each axis, and the moment of inertia $I$ at each step, which is calculated using}
\begin{equation}\label{eq:MOI}
    I=\int_S\rho\,\,\bigg(\,\frac{\|\boldsymbol{r}\times\boldsymbol{\Omega}\|^2}{\|\boldsymbol{\Omega}\|^2}\ \bigg)\, \text{d}x\,.
\end{equation}
{\texttt{SAMUS} is further capable of incorporating a broad class of user-defined functions in these outputs.}
 
 \subsection{{Trajectory Jump Method}}\label{subsec:trajjump}
 
{\texttt{SAMUS} uses a ``trajectory jump" method for efficiency, reducing the number of computations necessary over the trajectory. \texttt{SAMUS} first computes the time-averaged deformation over a (user-defined) number of rotational periods. It then performs a linear extrapolation of this average and steps forward in simulation time until either (i) the CFL condition is violated or (ii) the heliocentric distance changes by 1\% (this threshold is similarly user-defined). This method assumes that the rotational period of the body is significantly shorter than the timescale within which the object moves through its trajectory significantly. The quality of this first-order linear approximation of the distortion was validated with convergence tests, using halved tolerances (available in the package). However, for bodies with slower rotation, further testing should be performed to confirm the validity of this methodology.}
 
 \subsection{{Tidal Force Computation}}\label{subsec:SAMUStidalforce}
 
 In {\texttt{SAMUS}, \texttt{quaternion} is used to rotate points to the stationary frame, a necessary step to compute the the continuum tidal force over $\mathcal{S}_3$. Quaternions allow for rapid computation of rotation by an arbitrary angle about an arbitrary axis, without gimbal locking \citep{kuipers2007}. For a given point $\boldsymbol{r}$ and quaternion $\boldsymbol{q}$, the rotated point is $\boldsymbol{r}'=\boldsymbol{q}\boldsymbol{r}\boldsymbol{q}^*$. A rotation by an angle $\theta$ about an axis $\boldsymbol{\hat{\Omega}}=(\Omega_x,\Omega_y,\Omega_z)$ is described by $\boldsymbol{q}=\cos{(\theta/2)}+\Omega_x\sin{(\theta/2)}\boldsymbol{i}+\Omega_y\sin{(\theta/2)}\boldsymbol{j}+\Omega_z\sin{(\theta/2)}\boldsymbol{k}$ and conjugate $\boldsymbol{q^*}=\cos{(\theta/2)}-\Omega_x\sin{(\theta/2)}\boldsymbol{i}-\Omega_y\sin{(\theta/2)}\boldsymbol{j}-\Omega_z\sin{(\theta/2)}\boldsymbol{k}$. Here, $\boldsymbol{i}$, $\boldsymbol{j}$, and $\boldsymbol{k}$ are the imaginary unit quaternions. }

 {The position quaternion $\boldsymbol{r}'=x'\boldsymbol{i}+y'\boldsymbol{j}+z'\boldsymbol{k}$ in the co-rotating frame is given by $\boldsymbol{r}=\boldsymbol{q^*}\boldsymbol{x'}\boldsymbol{q}$ in the non-rotating frame. The tidal force at each point on the body is then computed by \texttt{SAMUS} using 
\begin{equation}\label{eq:samustide}
    \boldsymbol{F}_{\text{tidal}}(\boldsymbol{r}')=-\boldsymbol{\hat{x}}GM_{\Sun}\rho\left(\frac{1}{(r_H-(\boldsymbol{q^*}\boldsymbol{r}'\boldsymbol{q})_x)^2}-\frac{1}{r_H^2}\right)\,,
\end{equation}
for an object with constant density $\rho$.} 
 
{A pseudocode describing the algorithm used in \texttt{SAMUS} is given in Appendix \ref{sec:pseudocode}.}

\section{{Analytic Approximation of Tidal Deformation}}\label{sec:analytics}

{In this section, we present analytic estimates to approximate the tidal deformation. For a point mass at some distance $R$ from an object of mass $M$, the magnitude of the acceleration due to gravity is 
\begin{equation}\label{eq:acceleration_gravity}
    a_g = \frac{GM}{R^2}\,.
\end{equation}
In Equation \ref{eq:acceleration_gravity}, G is the gravitational constant. The difference in the acceleration across a distance $\Delta r$ is the tidal acceleration $a_{\rm tidal}$, which is given by
\begin{equation}
    a_{\rm tidal} = GM\,\bigg| \frac{1}{(R+\Delta r)^2}-\frac{1}{R^2}\bigg|\,.
\end{equation}
This simplifies to 
\begin{equation}
    a_{\rm tidal} = GM\bigg| \frac{\Delta r(\Delta r+2R)}{R^2(R+\Delta r)^2}\bigg|\,.
\end{equation}
 Assuming that $\Delta r \ll R$ and Taylor expanding about $\Delta r=0$ yields,
\begin{equation}\label{eq:tidalacc}
    a_{\rm tidal}\simeq 2GM\,\bigg|\frac{\Delta r}{R^3}\bigg|\,.
\end{equation}}

{Now let us consider the scale of deformation experienced by an object subjected to an external acceleration $a$. The force experienced by a continuum object of density $\rho$ is $a\rho$. The  magnitude of the tidal force per unit volume, $F_{\rm tidal}$, is 
\begin{equation}\label{eq:tidalforce}
    F_{\rm tidal}\simeq 2GM\rho \,\bigg|\frac{\Delta r}{R^3}\bigg|\,.
\end{equation}
For the dynamic shear viscosity $\mu$, we use the definition that $\sigma=2\mu\Dot{\epsilon}$, where $\sigma$ is the stress and $\Dot{\epsilon}$ is the time derivative of the strain. For a domain of length $L$, the strain $\epsilon$ is defined as the fractional change in the length of the domain, and its time derivative is simply $\Dot{\epsilon}=\Dot{L}(t)/L_0$, where $L_0$ is the initial domain length. On the other hand, the stress is the force per unit area, so we will write $\sigma = F_{\rm tidal} L_0$ with $L_0$ the characteristic length. }

\begin{figure*}
\centering
\includegraphics[width=\linewidth,angle=0]{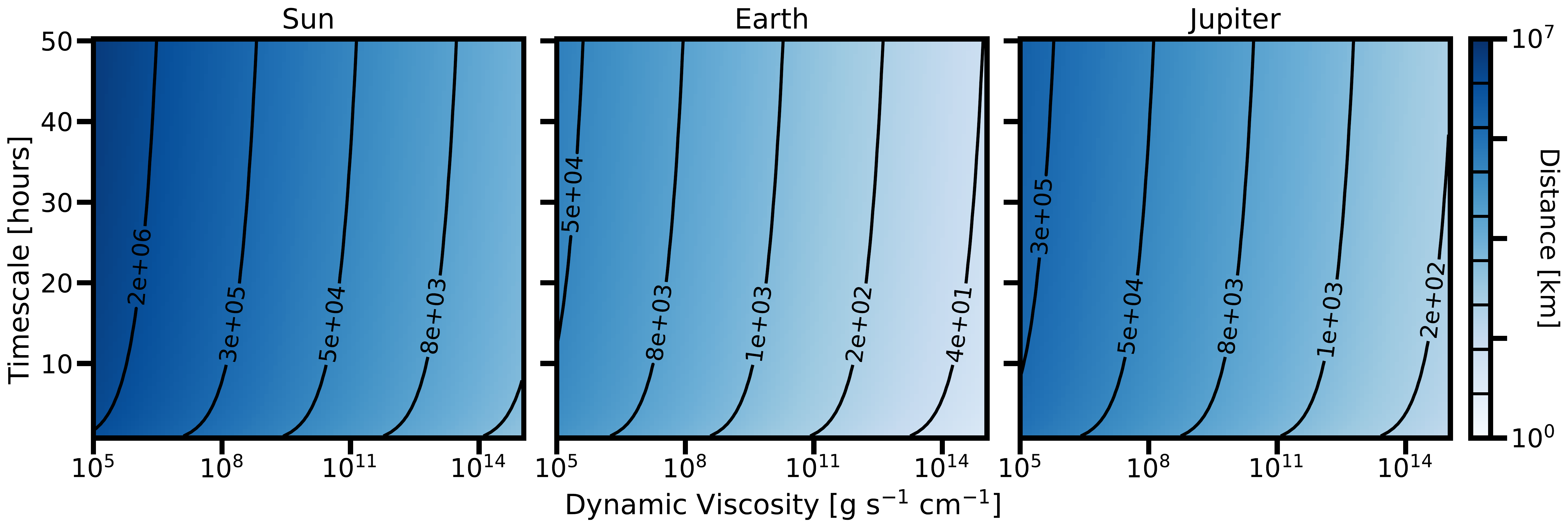}
\caption{ Distance at which an object of 100 m is deformed by 30\%, versus dynamic viscosity (in poise) and interaction timescale (in hours). Results are presented for cases in which the Sun, the Earth, and Jupiter are the primary orbital bodies. }
\label{fig:deformationanalysis}
\end{figure*}

{Therefore, we write}
\begin{equation}
    \underbrace{\frac{2GM\,\rho \Delta r}{R^3}}_\sigma = \underbrace{\frac{2\mu}{L_0}\frac{dL(t)}{dt}}_{2\mu\Dot{\epsilon}}\,.
\end{equation}
{In this equation, the left-hand side is $\sigma$, the force times a characteristic length, and the right-hand side is $2\mu\Dot{\epsilon}$. This approximate equation can be  solved, using Equation \ref{eq:tidalforce} and setting $\Delta r=L(t)$, since we wish to find the deformation of the entire object:}
\begin{equation}\label{eq:deformscaling}
   L(t)=L_0\, \exp{ \bigg[\frac{GM L_0^2\rho}{\mu R^3}t\bigg]}\,.
\end{equation}
{This equation allows us to compute the expected deformation for a given force and over a given timescale. This equation also exhibits the expected dependence --- increasing with time, density, and initial length and decreasing with the viscosity and the orbital distance. }

{Equation \ref{eq:deformscaling} allows for analysis of tidal deformation and material cohesion in general small bodies. In Figure \ref{fig:deformationanalysis}, we present heatmaps showing the distance at which an object will deform by 30\%, which we assume to be significant. The value of 30\% deformation is only an estimate, and a greater understanding of small-body cohesion is necessary to provide stricter constraints on structural collapse. We present these distances for objects of 100 m in radius, in orbit around the Sun, the Earth, and Jupiter for a range of dynamic viscosity and interaction timescales. These results may provide constraints on the cohesive stability of Sun-grazing comets and on tidally destructed comets such as Shoemaker-Levy 9. These results can also be used to constrain the cohesive properties of small bodies in the solar system in future research. }

\section{{Simplified Light Curve Model}}\label{sec:lightcurvefits}

{In this section, we {present a model} to generate synthetic light curves of minor bodies with the simplifying assumption of a {``}fixed-axis{"} NPA rotation. To validate our methodology, we adopt idealized rotational states and orbital geometries in Appendix \ref{sec:simplemodel}, which we compare against the more complex derived} values. We then describe a methodology to fit these synthetic data to photometric light curves to obtain parameters for a characteristic {{``}fixed-axis{"} NPA rotation for minor bodies. 

\subsection{{Light Curve Model}}

{In our light curve model, we incorporate the change in the phase angle due to the minor body's astrometric progression in its orbit, while restricting the rotation to follow a single {``}fixed-axis{"} NPA rotation. This is a stringent simplification, {as this sort of rotation is non-physical.} However, this model allows us to semi-accurately reproduce observed light curves and incorporate the simulated tidal deformation {for comparison}, balancing physical accuracy and practical restrictions.}

{In order to model the light curve, we} use Equation 10 from \citet{muinonen2015}, which we refer to as `ML15' for the remainder of this paper. We assume that {the relevant objects have} a diffuse Lommel-Seeliger scattering surface, which represents a closely-packed particulate medium with weak multiple scattering. Integrating the Lommel-Seeliger scattering {function} over the exposed-and-visible surface gives an expression for the brightness at any orientation.

We assume a single-scattering albedo $A$ and an isometric single-scattering phase function $P(\alpha)=1$. This incorporates modulation based on the body orientation, but assumes no additional modulation from the scattering function. We define the phase angle $\alpha$ as the interior Sun-object-Earth angle, and note that $\cos\alpha=\boldsymbol{\hat{e}}_{\Sun}\cdot\boldsymbol{\hat{e}}_{\Earth}$, where $\boldsymbol{\hat{e}}_{\Sun}$ and $\boldsymbol{\hat{e}}_\Earth$ are unit vectors in the direction of the Sun and Earth respectively. We also define the matrix $\boldsymbol{C}$ as
\begin{equation}
    \boldsymbol{C}\equiv\begin{pmatrix}
    a^{-2} & 0 & 0\\
    0 & b^{-2} & 0\\
    0 & 0 & c^{-2}
    \end{pmatrix}\,,
\end{equation}
and the parameters $T_{\Sun}$ and $T_\Earth$ to be
\begin{equation}
    \begin{aligned}
        T_{\Sun}\equiv&\sqrt{\boldsymbol{\hat{e}}_{\Sun}^TC\boldsymbol{\hat{e}}_\Sun}\\
        T_\Earth\equiv&\sqrt{\boldsymbol{\hat{e}}_\Earth^TC\boldsymbol{\hat{e}}_{\Earth}}\,.
    \end{aligned}
\end{equation}
We further define the parameter $T$ as
\begin{equation}
    T\equiv\sqrt{T_{\Sun}^2+T_\Earth^2+2T_{\Sun}T_\Earth\cos\alpha'}\,,
\end{equation}
the angle $\alpha'$ as 
\begin{equation}
\begin{aligned}
    \cos\alpha'=&\frac{\boldsymbol{\hat{e}}_{\Sun}^TC\boldsymbol{\hat{e}}_\Earth}{T_{\Sun}T_\Earth}\\
    \sin\alpha'=&\sqrt{1-\cos^2\alpha'}\,,
\end{aligned}
\end{equation}
and the angle $\lambda'$ as 
\begin{equation}
\begin{aligned}
    \cos\lambda'=&\frac{T_\Sun+T_\Earth\cos\alpha'}{T}\\
    \sin\lambda'=&\frac{T_\Sun\sin\alpha'}{T}\,.
\end{aligned}
\end{equation}
Then the disk-integrated brightness $L_{\rm ML15}$ is 
\begin{equation}
\begin{aligned}
    L_{{\rm ML15}}(\alpha)=&\frac{1}{8}\pi F_0AP(\alpha)abc\frac{T_\Sun T_\Earth}{T}\\
    &\bigg(\cos(\lambda'-\alpha')+\cos\lambda'+\sin\lambda'\sin(\lambda'-\alpha')\\
    &\ln{\big[\cot(\frac{1}{2}\lambda')\cot(\frac{1}{2}(\alpha'-\lambda'))\big]}\bigg),
\end{aligned}
\end{equation}
where $\pi F_0$ is the incident flux density. The absolute magnitude is given by
\begin{equation}\label{eq:MLmodel}
\begin{aligned}
        H=\Delta V-&2.5\log\bigg(abc\frac{T_\Sun T_\Earth}{T}\\
        &\big(\cos(\lambda'-\alpha')+\cos\lambda'+\sin\lambda'\sin(\lambda'-\alpha')\\
        &\ln{\big[\cot(\frac{1}{2}\lambda')\cot(\frac{1}{2}(\alpha'-\lambda'))\big]}\big)\bigg),
\end{aligned}
\end{equation}
where $\Delta V$ is a constant to absorb the flux and magnitude conversion \citep{mashchenko2019}.

\begin{figure}
\centering
\includegraphics[width=\linewidth,angle=0]{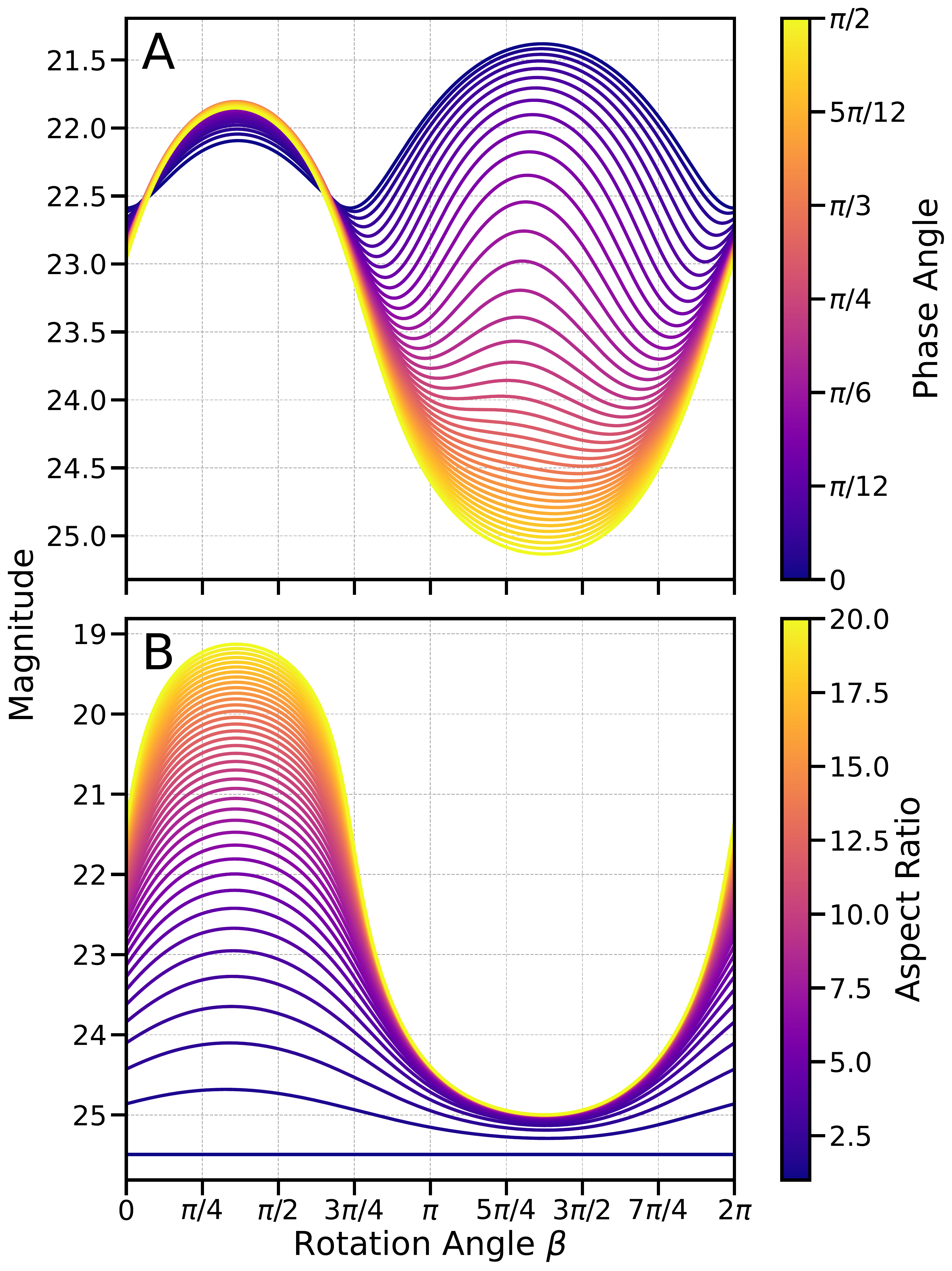}
\caption{Synthetic light curves of a body rotating with an arbitrary period for a range of phase angle (A) and aspect ratio (B). }
\label{fig:phaseeffects}
\end{figure}

\subsection{Phase Angle and Aspect Ratio Dependence}\label{subsec:phaseaspectanalysis}

In this subsection, we investigate the dependence of the ML15 model on phase angle and aspect ratio. {This will be applicable to `Oumuamua in Section 4.3.} We compute synthetic light curves for a single (arbitrary) period with varying phase angle and aspect ratio, and set the parameters ($\theta$, $\phi$, $\psi$, $\beta_0$, and $\Delta V$) to the optimized values given in Section  \ref{subsec:gettingaxis}. However, we verified that the phase angle-- and aspect ratio-- dependent behavior of the light curve does not sensitively depend on the other parameters. This analysis both validates the ML15 model and explains components of the synthetic light curves in Figure \ref{fig:optimalaxiscurvesimdata}. 

We present 40 synthetic light curves with phase angles ($\alpha$) uniformly distributed in [$0$,$\pi/2$) and with an aspect ratio of 6:6:1 in Figure \ref{fig:phaseeffects}A. For low phase angle, there are two distinct peaks at $\beta=\pi/2$ and $\beta=3\pi/2$, caused by the half-period cycles of illumination from the larger and smaller cross-sections. As $\alpha\rightarrow\pi/2$, the light curve becomes approximately sinusoidal, as at $\alpha=\pi/2$ only a single face of the body is observable, with brightness variation due to changing exposure. 

We also show 40 synthetic light curves for aspect ratios ranging from 1 to 20 and with $\alpha=\pi/2$ in Figure \ref{fig:phaseeffects}B. Here, the magnitude variation scales with increasing aspect ratio and is zero for a sphere. This effect is simply due to the increasing cross-sectional area contrast for larger aspect ratios. 

\subsection{Obtaining {a Characteristic} Rotation Axis}\label{subsec:gettingaxis}

In this subsection, we {describe the use of} the ML15 model to find {a characteristic rotational state for an arbitrary body} under the assumption of a {``}fixed-axis{"} NPA rotation. To find this rotational state, we fit a synthetic light curve generated with Equation \ref{eq:MLmodel} to {photometric data (assumed to be} corrected for light travel time, helio- and geo-centric distance, {and Solar magnitude}). We must carefully note that a {``}fixed-axis{"} NPA rotation is non-physical, and does not exist in nature. As such, this model cannot be used to describe the physical rotation of an object, but only to provide a well-fitting light curve which can be compared to a deformed model.

We use the SciPy package's \texttt{scipy.optimize.curve\_fit} for the optimization, which uses a non-linear least squares algorithm to minimize $\chi^2$, where
\begin{equation}
\begin{aligned}
    \chi^2\equiv&\sum_i\left(\frac{(y_i-\mu_i)^2}{\sigma_{y_i}^2}\right)\,.
\end{aligned}
\end{equation}
Here, $y_i$ and $\sigma_{y_i}$ denote photometric measurements and associated errors, while $\mu_i$ denotes corresponding synthetic values. 

{The parameters that we optimize} define the rotation axis and the Earth-pointing axis. The Sun-pointing direction $\boldsymbol{\hat{e}}_\Sun$ is {fixed to be} along the $\boldsymbol{\hat{x}}$ direction, {reducing symmetric degeneracy.} For a {given} phase angle $\alpha$, the direction of the observer $\boldsymbol{\hat{e}}_\Earth$ is therefore constrained to a cone centered on the x-axis such that $\boldsymbol{\hat{e}}_\Sun\cdot\boldsymbol{\hat{e}}_\Earth=\boldsymbol{\hat{x}}\cdot\boldsymbol{\hat{e}}_\Earth=\cos\alpha$. We then define a new variable $\theta$ such that the direction of the observation is $\boldsymbol{\hat{e}}_\Earth=\cos\alpha\,\boldsymbol{\hat{x}}+\sin\alpha\cos\theta\,\boldsymbol{\hat{y}}+\sin\alpha\sin\theta\,\boldsymbol{\hat{z}}$, with $\theta=0$ if $\boldsymbol{\hat{e}}_\Earth$ lies in the x-y plane.

We additionally parameterize the rotation axis $\boldsymbol{\hat{\Omega}}$ by spherical coordinates about the x-axis. We use a polar angle $\phi$ such that $\boldsymbol{\hat{\Omega}}\cdot\boldsymbol{\hat{x}}=\cos\phi$ and an azimuthal angle $\psi$ measured from the y-axis and restricted to the y-z plane. With these definitions, $\boldsymbol{\hat{\Omega}}=\cos\phi\,\boldsymbol{\hat{x}}+\sin\phi\cos\psi\,\boldsymbol{\hat{y}}+\sin\phi\sin\psi\,\boldsymbol{\hat{z}}$. We restrict the optimization to $\theta\in[0,2\pi)$, $\phi\in[0,\pi/2]$, and $\psi\in[0,2\pi)$ to reflect both the modular domain and symmetry. We optimize two additional variables from Equation \ref{eq:MLmodel} --- the initial attitude about the rotation axis $\beta_0$, and the constant $\Delta V$, which parameterizes the flux and albedo. We also optimize the rotational period $p$, which we restrict to {sufficiently encompass} all previously measured values. This period maps the time of observation $t$ to the rotation attitude $\beta$ via $\beta=2\pi\cdot(t\%p)/p$, a continuous linear mapping from $(t\in[0,\infty))\rightarrow(\beta\in[0,2\pi))$. Validation of the search method {via fitting to simple rotators} is presented in Section \ref{sec:lightfitvalid}.

The parameter space has many minima because it is highly degenerate and interdependent --- (i) there are multiple well-fitting {``}fixed-axis{"} rotations due to {tumbling in these objects, }(ii) there is often weak dependence of the light curve on parameter sets, and (iii) there are distinct parameter sets which produce relatively similar light curves {due to the highly symmetric equations.} Therefore, the optimal parameters depend strongly on the initial conditions. To account for this, we perform a {grid-search optimization} over this data set, using {a set of }grid-spaced initial {values} for the parameters ($p$, $\theta$, $\phi$, $\psi$, $\beta_0$, and $\Delta V$). {For each point in the grid, \texttt{curve\_fit} is used to obtain a parameter set that describes a local minimum in the parameter space. The final optimized parameters are} the parameter set with the lowest $\chi^2$ {out of all local minima found with the initial parameter values. }

{Note that this model does not incorporate the effects of a changing period, tumbling, or torques, limiting its applicability for objects with complex rotational states. However, this model is sufficiently accurate for use in the \texttt{SAMUS} software, due to the limited effects of the rotational state on the magnitude of the tidal deformation. }

\begin{figure}
  \centering
  \begin{minipage}[b]{0.45\textwidth}
    \includegraphics[width=\textwidth]{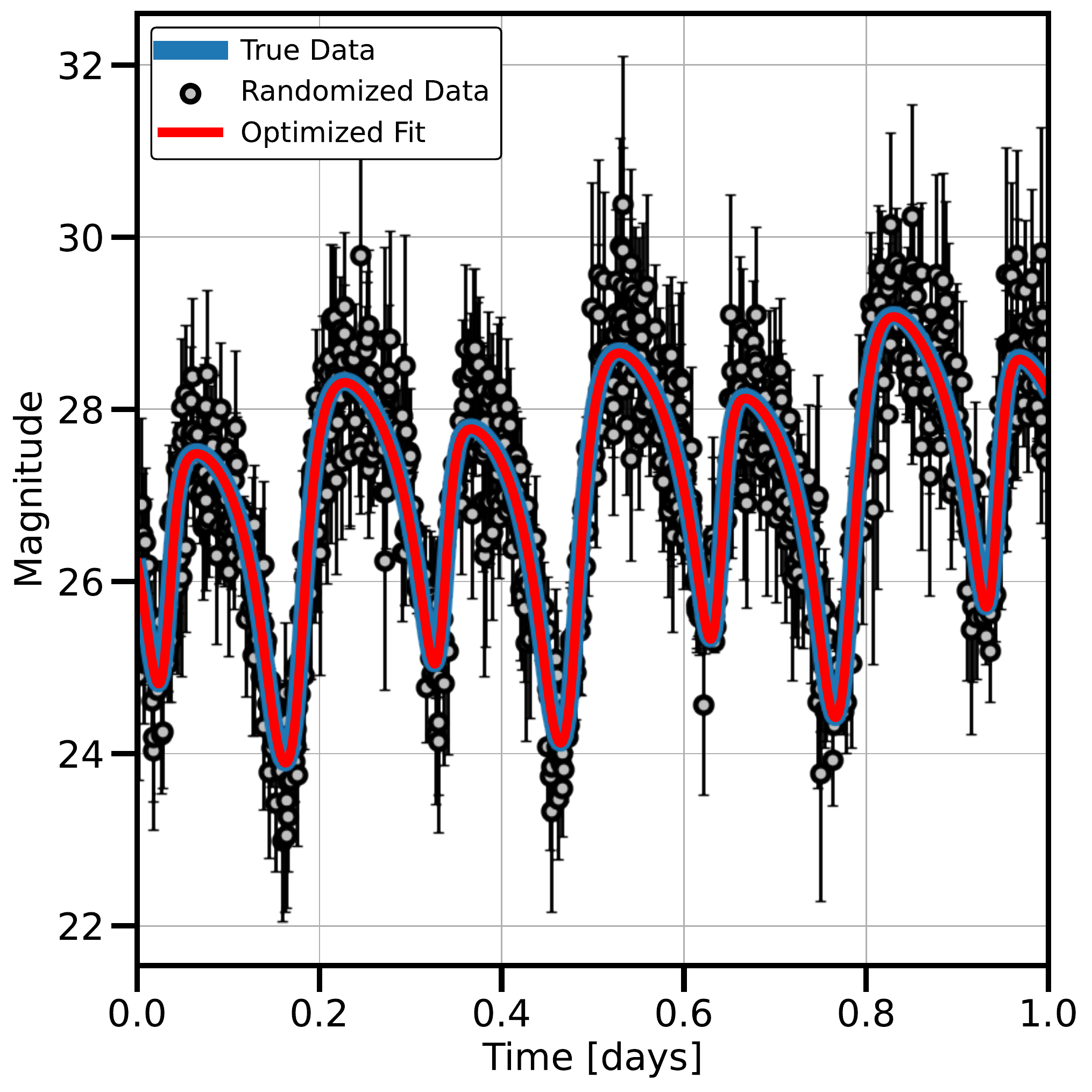}
    \caption{{Optimized fit (red line) for a random parameter choice. The true (blue line) and noise-added data (grey points) are also plotted for comparison. }}
    \label{fig:light_valid}
  \end{minipage}
  \hfill
  \begin{minipage}[b]{0.45\textwidth}
    \includegraphics[width=\textwidth]{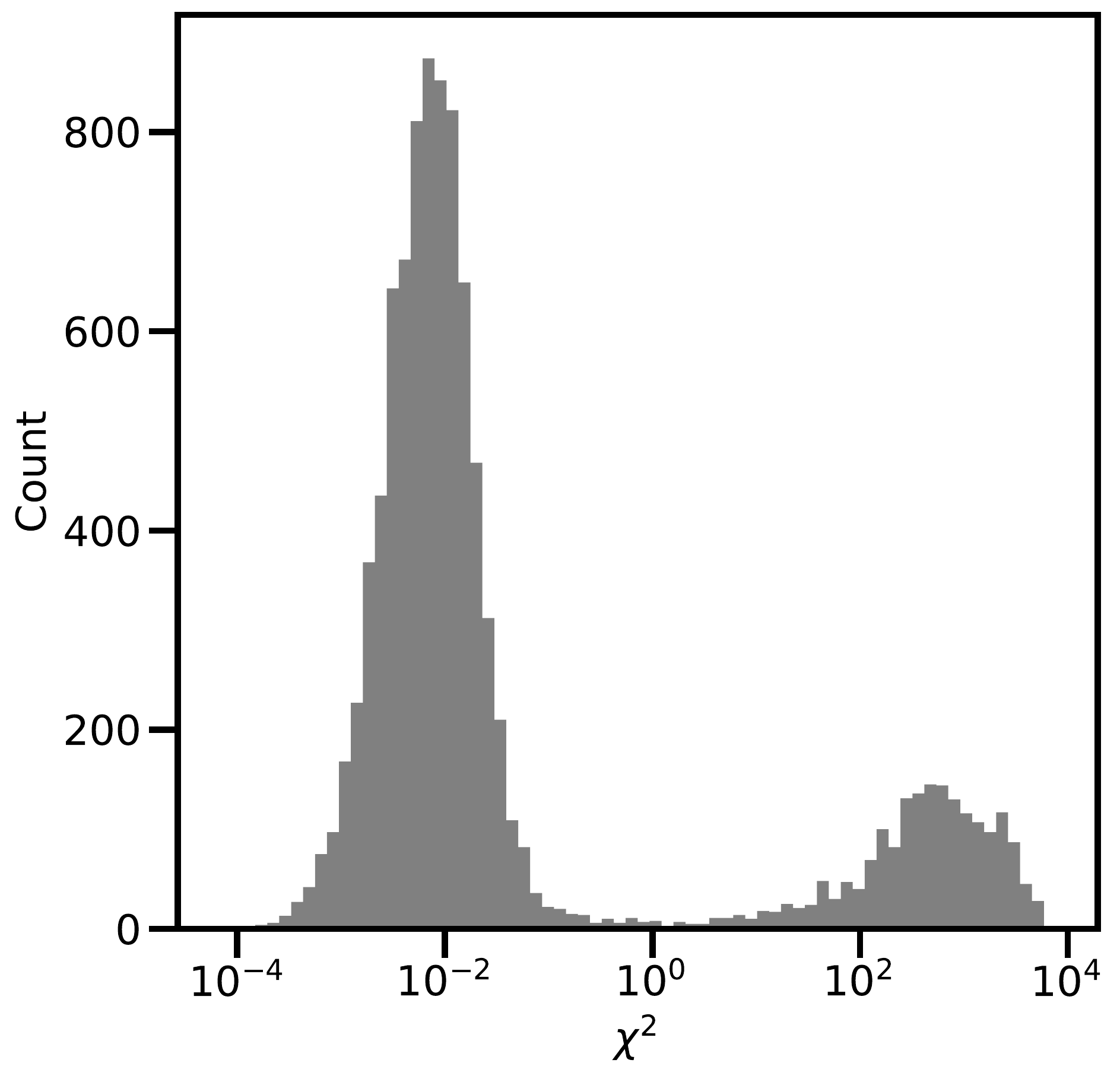}
    \caption{{The distribution of $\chi^2$ for $10^4$ synthetic light curves. The histogram marginalizes over all selected parameters, which were randomly chosen. }}
    \label{fig:chi2_valid}
  \end{minipage}
\end{figure}

\subsection{{Validation with Simulated Simple Rotators}}\label{sec:lightfitvalid}

{In this subsection, we validate our use of \texttt{scipy.optimize.curve\_fit} to optimize the rotation axis by sitting a series of simple rotators. We generate random combinations of parameters $p$, $\theta$, $\phi$, $\psi$, $\beta_0$, and $\Delta V$ which are used to generate synthetic light curves for a 1-day period {(using the ML15 model)}. We add Gaussian random errors to the synthetic data, with standard deviations three times the mean of the errors in the {`Oumuamua} photometric data. We also re-scaled the error by the log of the square root of the magnitude, to imitate photon noise.}

{We fit the ML15 model to the randomized data using random initial conditions drawn from Gaussian distributions about the true values. These initial conditions are unrealistically close to the true values compared to what we can achieve when optimizing real data. However, the {grid-search} optimization performed in Section \ref{subsec:gettingaxis} overcomes this drawback and produces {high-quality} fits.}

{In Figure \ref{fig:light_valid}, we show the true and noise added data and the optimal fit for a randomly drawn parameter set. This fit is extremely high quality, and the true value, optimized value, and statistical difference of the parameters are given in Table \ref{table:lightvalid}. Testing a variety of random parameters provides similar results (not shown). }

\begin{table}
\centering
\begin{tabular}{|c|c|c|c|}
 \hline
 \multicolumn{4}{||c||}{Parameters} \\ \hline\hline
 Parameter & True Value & Fit Value & Stat. Diff. \\ \hline
 $p$ & 7.2952 & 7.2954 & 0.414 \\ \hline
 $\theta$ & 5.7832 & 5.7832 & 0.022 \\ \hline
 $\phi$ & 0.5935 & 0.5916 & 1.418 \\ \hline
 $\psi$ & 0.2803 & 0.2797 & 1.043 \\ \hline
 $\beta_0$ & 0.3523 & 0.3522 & 0.252 \\ \hline
 $\Delta V$ & 33.3230 & 33.3265 & 1.402 \\ \hline
\end{tabular}
\caption{True and derived parameters for the synthetic light curve shown in Figure \ref{fig:light_valid}.}
\label{table:lightvalid}
\end{table}

{We also perform this test for $10^4$ randomly sampled values, and plot their accuracy versus the estimated values in Figure \ref{fig:chi2_valid}. The majority of fits to synthetic data are very high-quality with $\chi^2\simeq10^{-2}$. Importantly, we verified that the distribution of $\chi^2$ shown in this figure is independent of the value of the randomized parameters and of the parameter itself (not shown for the sake of brevity). Clusters close to $\chi^2\simeq10^{-2}$ are nearly perfect fits, similar to those presented in Figure \ref{fig:light_valid}. This indicates that this optimization method is effective for any underlying set of parameters. }

\section{{Applicability of \texttt{SAMUS} to 99942 Apophis}}

{As discussed in the introduction, the near-Earth object (NEO) 99942 Apophis will make a close approach to the Earth in 2029 \citep{Giorgini2008,Farnocchia2013}. Nongravitational perturbations from the Yarkovsky effect could produce an impact event in the 2068 close approach with a probability $>10^{-6}$ \citep{Farnocchia2013}. The 2029 close approach will expose the object to significant tidal forces that most likely will not produce catastrophic disruption but could produce local failures \citep{Scheeres2005,Yu2014}. The strength of such forces makes Apophis an excellent candidate for further investigation with \texttt{SAMUS}. While large-scale deformation is likely negligible \citep{Yu2014}, the magnitude (or absence) of deformation in Apophis will provide information on its material properties. Notably, in 2029, caution must be taken to correct for predicted variations in the light curve due to tidal resurfacing \citep{Yu2014,Binzel2010,Kim2023} and alterations in Apophis' spin state \citep{Scheeres2005,Benson2023}. In this section, we discuss the potential effects of this close approach on Apophis' shape. }

\begin{figure}
    \centering
    \includegraphics[width=\linewidth,angle=0]{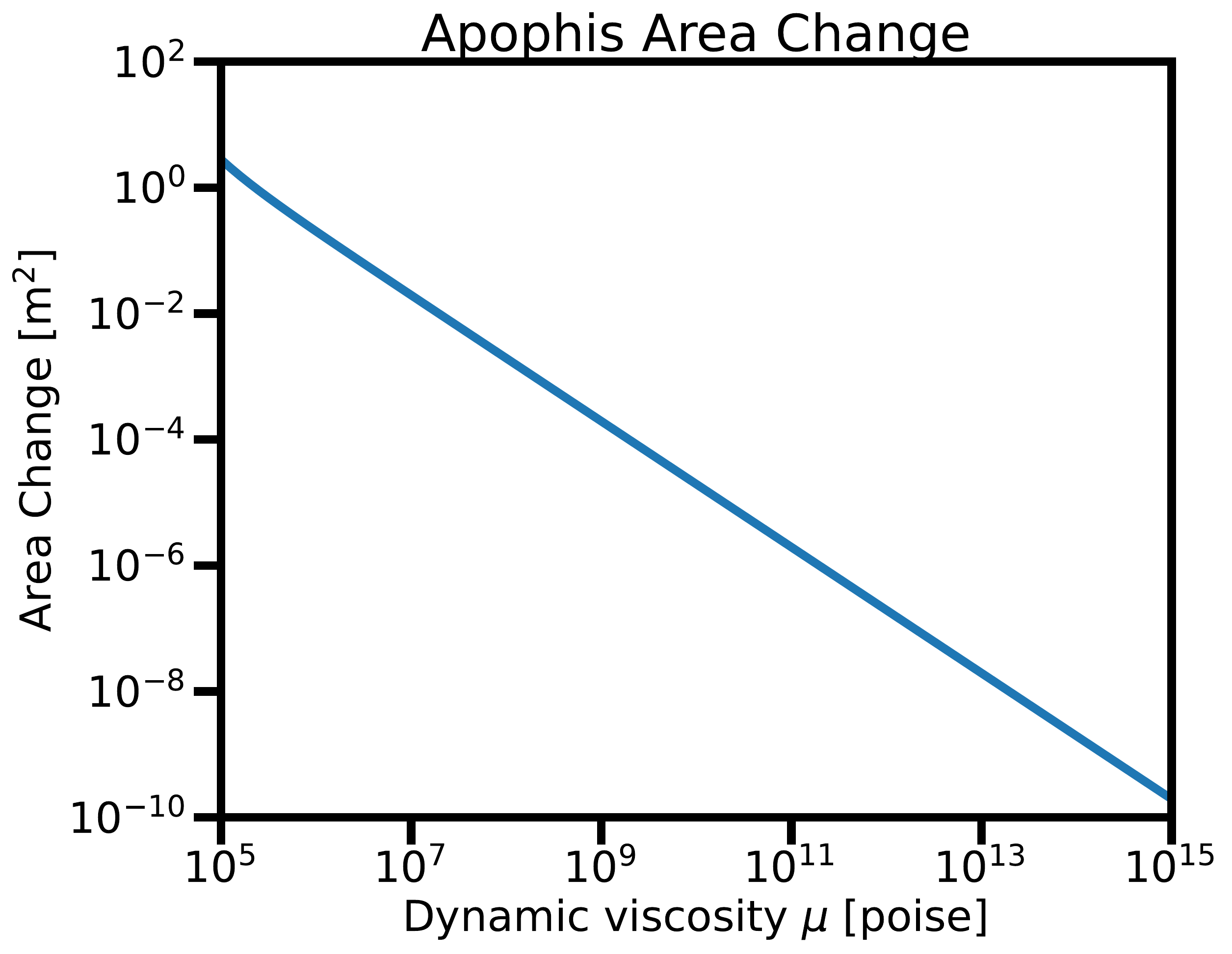}
    \caption{Order-of-magnitude area change of Apophis for a range of dynamic viscosities, in m$^2$.}
    \label{fig:apophis_area}
\end{figure}

{At its closest approach, Apophis will be within 38,000 km of the Earth \citep{Farnocchia2013}. Using Equation \ref{eq:tidalacc}, we find that Apophis will experience a maximum tidal acceleration of $4.14\times 10^{-4}$ cm s$^{-2}$, using $M=5.97\cdot10^{27}$ g and $\Delta r=170$ m \citep{Lee2022}. }

{On closest approach, Apophis will spend approximately 4 hours closer than 50,000 km (\url{https://ssd.jpl.nasa.gov/horizons.cgi}). Using the tidal acceleration computed above, we can use Equation \ref{eq:deformscaling} to compute the expected deformation. The deformation for a range of viscosities is given by Figure \ref{fig:apophis_area} assuming a density of $\rho=1.5$ g cm$^{-3}$, and a radius of $L_0\simeq 170$ m \citep{Lee2022}. While high viscosities show little deformation in the object, detectable changes are likely possible for a range of low viscosities. For example, a  dynamic viscosity of $\mu\simeq10^{7}$ poise yields a deformation of $L\sim 1$ m. This will increase the surface area of Apophis by approximately $0.5\%$, assuming an ellipsoidal cross section with the vertical axes. Therefore, low dynamic viscosities could create potentially detectable effects on the photometric magnitude. However, larger viscosities will not produce detectable signals.  }

{\texttt{SAMUS} will then be a useful tool to constrain the material properties of Apophis, using data from the closest approach. This analysis will be especially relevant in combination with the OSIRIS-APEX mission, which will encounter Apophis during the asteroid's 2029 flyby \citep{Nolan2021}. The \textit{in situ} measurements, in combination with remote observations, will be useful for the analysis of Apophis' composition. The use of \texttt{SAMUS} would be especially powerful in combination with the analysis of \citet{Hirabayashi2022}, who demonstrated that changes in the rotational state are potential probes of Apophis' material properties.}

{However, Apophis' rotation state is currently in the process of non-principal axis (NPA) rotation and is slowly tumbling \citep{Pravec2014,Benson2023}. Due to the difficulties of interpreting data from this sort of light curve, we must develop a sophisticated NPA rotation model for incorporation into \texttt{SAMUS}. This rotation model will also be invaluable in accounting for changes in the rotation state due to gravitationally-induced torques \citep{Scheeres2005,Benson2023}.}

\begin{figure*}
\centering
\includegraphics[width=\textwidth,angle=0]{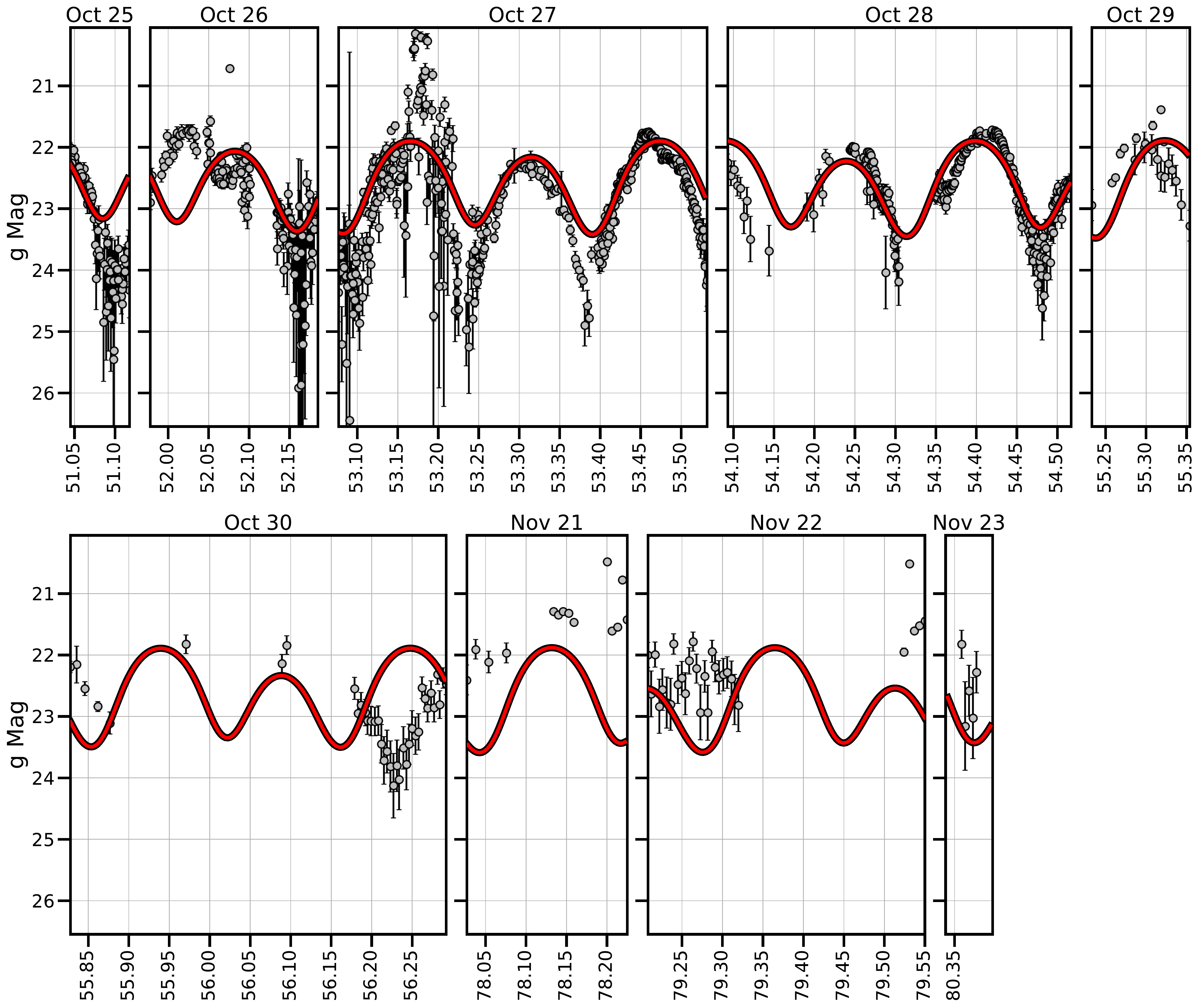}
\caption{{Synthetic light curve using the ML15 model (red line) alongside photometric observations of `Oumuamua (grey points, data from \citet{belton2018}). The period, initial rotation state, and average magnitude are optimized for the first 6 nights. }}
\label{fig:evolvinglightcurve}
\end{figure*}

\begin{figure}
\centering
\includegraphics[width=\linewidth,angle=0]{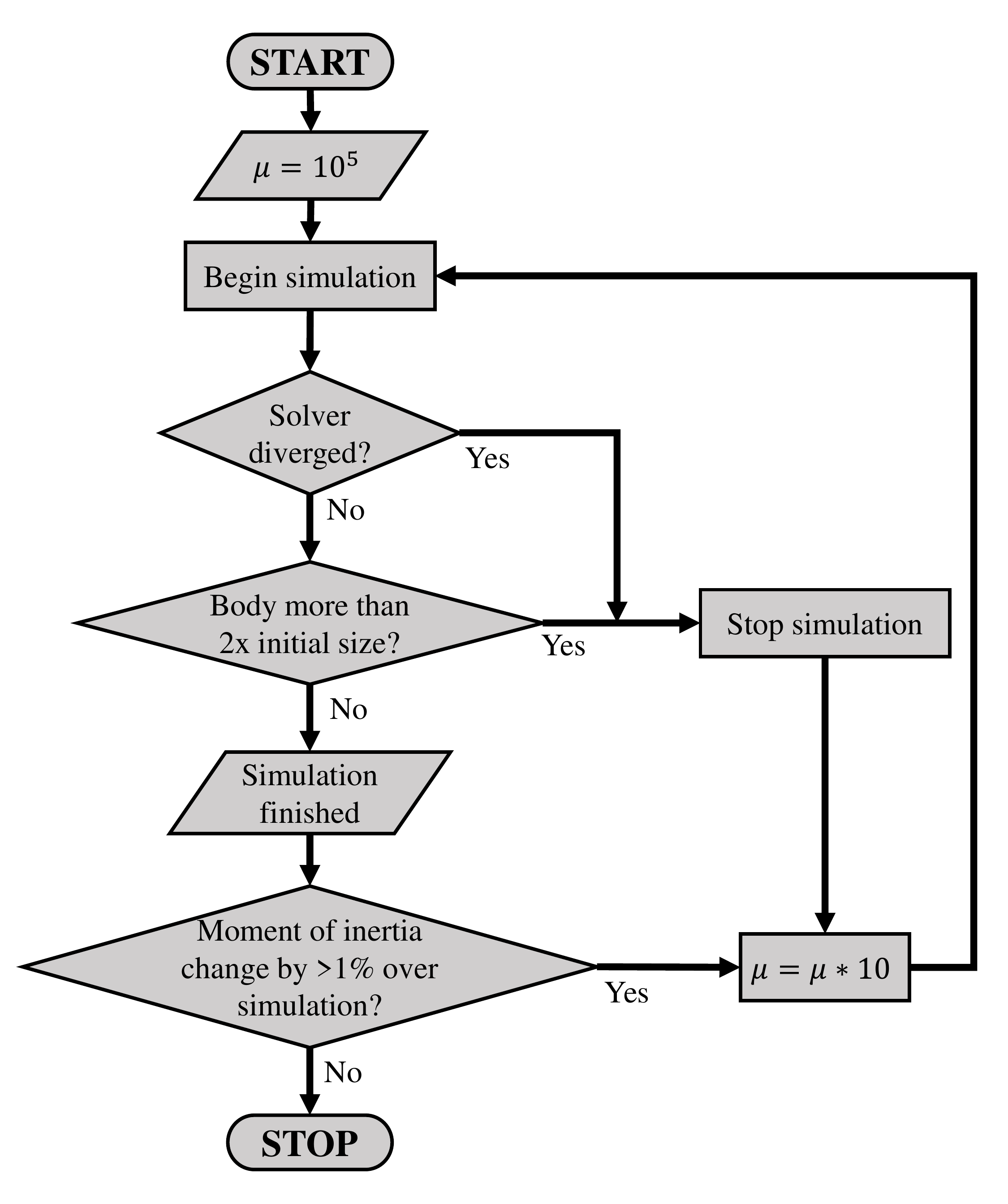}
\caption{Flowchart showing the structure of the simulation runs used to model `Oumuamua.}
\label{fig:simstruct}
\end{figure}

\begin{figure}
\centering
\includegraphics[width=\linewidth,angle=0]{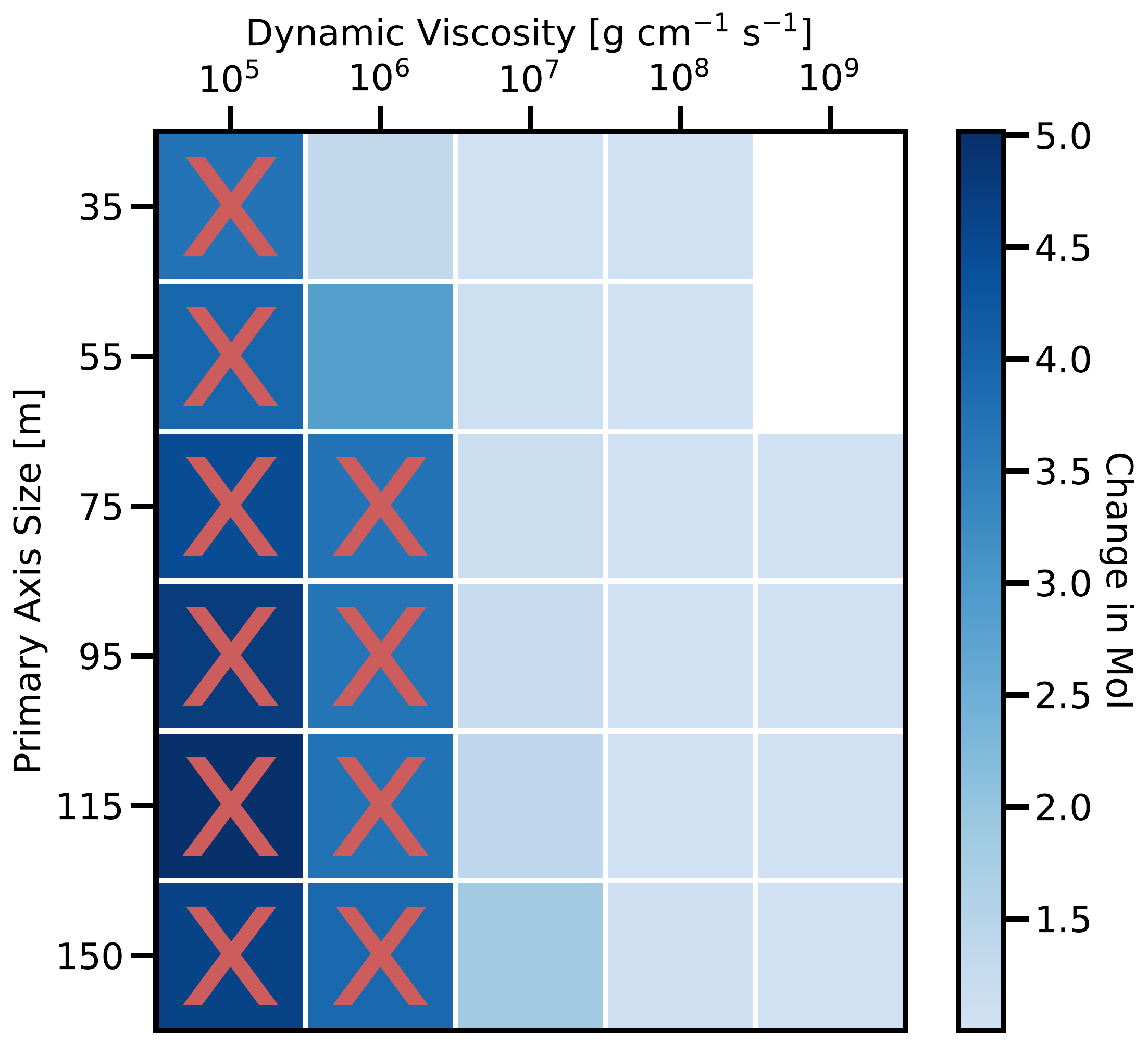}
\caption{Simulated change of `Oumuamua's moment of inertia from 2017 May 4 to 2018 January 16. Red X's indicate simulations that were halted because of disintegration or numerical divergence. Empty squares indicate parameters for simulations that were not run, as convergence was achieved for lower dynamic viscosity values. }
\label{fig:optimalaxisheatmap}
\end{figure}

\section{{`Oumuamua Simulation}}\label{sec:oumuamuasim}

In this section, we {use} \texttt{SAMUS} to constrain the {size and dynamic viscosity }of `Oumuamua{, demonstrating the effectiveness of this software for investigating objects which are potential candidates for tidal deformation. We first use the methodology described in Section \ref{sec:lightcurvefits} to obtain a characteristic {``}fixed-axis{"} NPA rotation for `Oumuamua, which we then incorporate into a \texttt{SAMUS} model for the simulation of tidal deformation. {It is important to note that the rotation model found for `Oumuamua is non-physical. Properly modeling NPA rotation is necessary for a full understanding of the case of `Oumuamua, which we leave for future research.} In these models, we adopt an aspect ratio of 6:6:1, following \citet{mashchenko2019}. We also make the approximation that the granular constituents of `Oumuamua are sufficiently small, such that the bulk aggregate behaves as a fluid.} 

\subsection{{`Oumuamua Axis Fitting}}\label{subsec:oumuamuaaxis}

{In order to find the characteristic rotation axis, we use the methodology described above to fit a {``}fixed-axis{"} NPA rotation to the photometric data collected in \citet{belton2018}. These data are corrected for light travel time, helio- and geo-centric distance, Solar magnitude, and filter color \citep{belton2018}, and the phase angle data are taken from JPL's Horizons database (}\url{https://ssd.jpl.nasa.gov/horizons.cgi}{). The optimal parameters (from a grid search over 3,000 points) are given in Table \ref{table:optimalparams}.} {Although this model does not accurately capture the physics of `Oumuamua's rotation, it provides a basic light curve for comparison to modeled deformation.}

\begin{table}[ht]
\begin{tabular}{ |c|r| } 
 \hline
 \multicolumn{2}{||c||}{OPTIMAL PARAMETERS} \\
 \hline\hline
 Parameter: & \multicolumn{1}{c|}{Value:} \\ 
 \hline
 $p$ [h] & $7.3975\pm1.11\cdot10^{-3}$ \\
 \hline
 $\theta$ & $4.6995\pm1.69\cdot10^{-2}$ \\
 \hline
 $\phi$ & $1.4193\pm1.32\cdot10^{-3}$ \\
 \hline
 $\psi$ & $1.2545\pm3.77\cdot10^{-3}$ \\
 \hline
 $\beta_0$ & $4.3950\pm9.82\cdot10^{-3}$ \\
 \hline
 $\Delta V$ & $32.1560\pm1.26\cdot10^{-2}$ \\
 \hline
\end{tabular}
\caption{The optimal parameters for the ML15 model in comparison to the photometric data. {Note that the errors here are not for the grid-search optimization but for the single, maximal optimization, as the grid-search uncertainty is unknown, and obtaining the uncertainty through Markov Chain Monte Carlo methods is computationally infeasible for this problem.}}
\label{table:optimalparams}
\end{table}

{These parameters correspond to a rotation axis of $\boldsymbol{\hat{\Omega}} =0.1509\boldsymbol{\hat{x}}+0.3075\boldsymbol{\hat{y}}+0.9395\boldsymbol{\hat{z}}$. This is relatively consistent with physical expectations, which suggests that the rotation axis should approach the smallest principal axis over time due to relative rotational energies. }

{We show the synthetic light curve for these optimal parameters in Figure \ref{fig:evolvinglightcurve}. The fit is qualitatively accurate for the October nights, but matches the November data poorly. This effect may be a result of a secular change in the spin period. However, proper evaluation of this effect requires the computation of outgassing-induced rotational dynamics, which are addressed in an accompanying paper. While the lack of tumbling in the model causes the fit to be non-exact, even in October, this does not have a significant effect on the tidal deformation effects addressed in this paper.}

\subsection{{`Oumuamua Tidal Deformation}}

{We adopt an aspect ratio of 6:6:1} \citep{mashchenko2019} and a bulk density of $\rho=0.5$ g cm$^{-3}$, which is typical for Solar System comets \citep{britt2006}. The viscosity is a parameter in these experiments, but is assumed to be {homogeneous within `Oumuamua.} We run simulations with an initial primary axis of $a=35,55,75,95,115$, and 150 meters in diameter, using the {characteristic {``}fixed-axis{"} NPA rotation described in Section {\ref{subsec:oumuamuaaxis}}. The rotation axis is $\boldsymbol{\hat{\Omega}}=0.1509\boldsymbol{\hat{x}}+0.3075\boldsymbol{\hat{y}}+0.9395\boldsymbol{\hat{z}}$, and we assume a constant rotational period of $p=7.3975$ hours. While these are held constant in each simulation, if the simulated body undergoes significant deformation, {conservation of angular momentum will produce} a change in the spin period. The simulations are run from 2017 May 4 to 2018 January 16, using trajectory data obtained from the \href{https://ssd.jpl.nasa.gov/horizons.cgi}{JPL Horizons database}.

We also ran simulations with rotations about each principal axis, and verified that this does not significantly affect the results, validating our use of the {characteristic rotation axis and our conclusions.} These simulations are not shown, {but their results are available on \href{https://github.com/astertaylor/Oumuamua}{GitHub}.}

While the dynamic viscosity depends on the temperature, neither are fully known for `Oumuamua. Therefore, we simplify the problem by performing these numerical experiments for a variety of effective viscosities, ignoring the temperature dependence. We initialize each simulation with a dynamic viscosity of $\mu=10^5$ g cm$^{-1}$ s$^{1}$, approximately equal to that of peanut butter under high pressure \citep{citerne2001}. The simulation is halted if the body is distorted to more than twice its initial size, as the cometary materials should disintegrate when subjected to such significant force ({although the shear tolerances of cometary materials are unknown}). The simulation is reset and the viscosity increased by an order of magnitude if (i) the solvers for the Navier-Stokes equations fail to converge, {indicating a divergence beyond physical conditions,} (ii) the {simulation was halted due to a non-physical size increase, or (iii) the} moment of inertia changes by more than 1\% over the path. These simulations are run with a trajectory jump tolerance of 1\% (see Section  \ref{subsec:trajjump}), and $C_{\text{max}}=1$. We compute 10 time steps over each rotational period, which fully samples the force over the rotation. For clarity, a flowchart describing this structure is shown in Figure \ref{fig:simstruct}, and the code used to create these simulations is available on \href{https://github.com/astertaylor/Oumuamua}{GitHub}.

\begin{figure}
\centering
\includegraphics[width=\linewidth,angle=0]{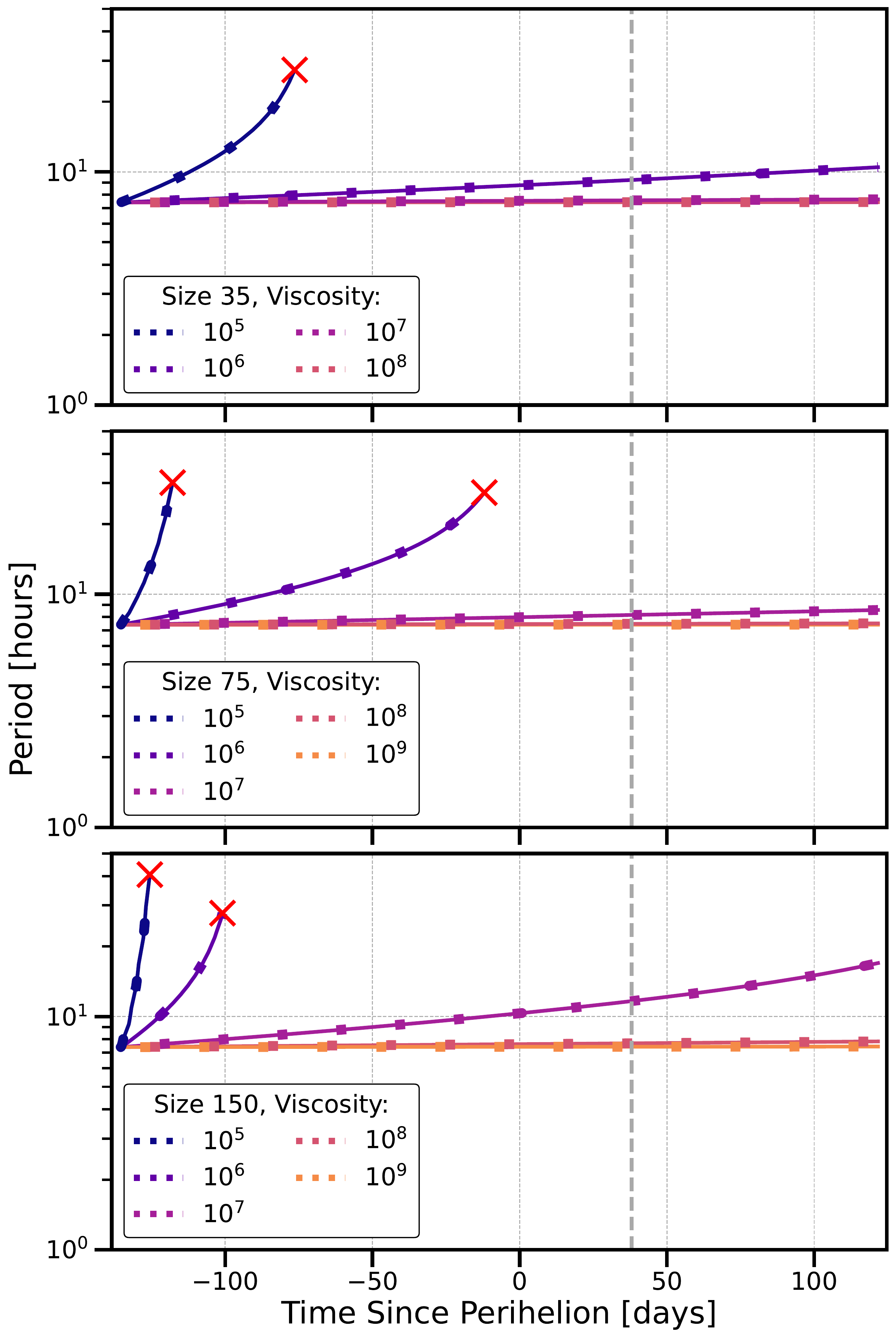}
\caption{Evolution of `Oumuamua's spin period due to modulation in the moment of inertia. The period is initialized at $p=7.3906$, ensuring that $p=7.3975$ hours at discovery (vertical dashed line) for the optimal fit. Rows represent different initial sizes. }
\label{fig:optimalaxisperiods}
\end{figure}

\begin{figure*}
\centering
\includegraphics[width=\linewidth,angle=0]{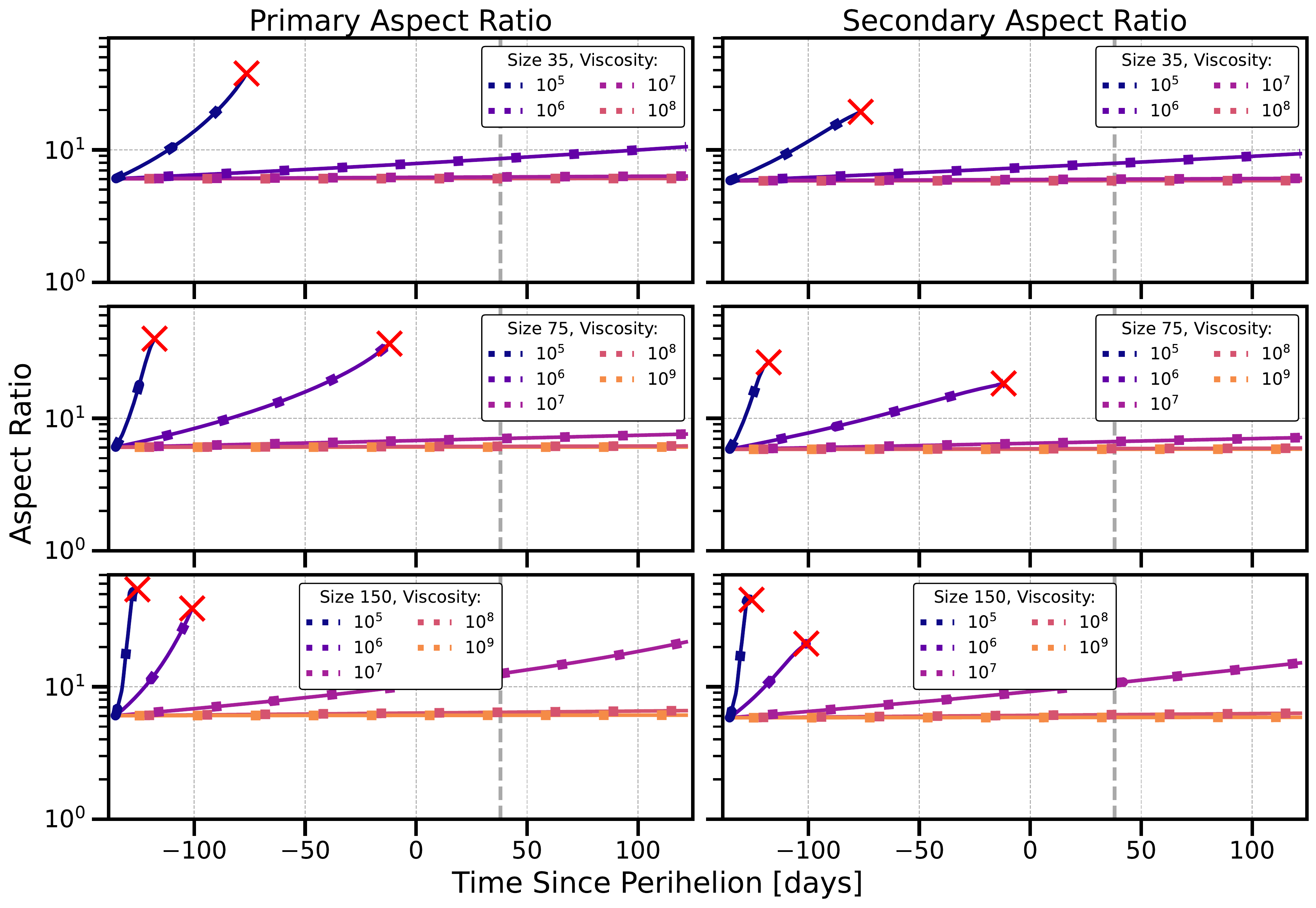}
\caption{Evolution of `Oumuamua's aspect ratios due to tidal deformation. The primary aspect ratio (left column) is the ratio of the largest to the smallest axis, while the secondary (right column) is the ratio of the intermediate to the smallest axis. At $t=0$, both aspect ratios are 6. The time of discovery is plotted as a vertical dashed line.}
\label{fig:optimalaxisaspectratio}
\end{figure*}

\begin{figure*}
\centering
\includegraphics[width=\linewidth,angle=0]{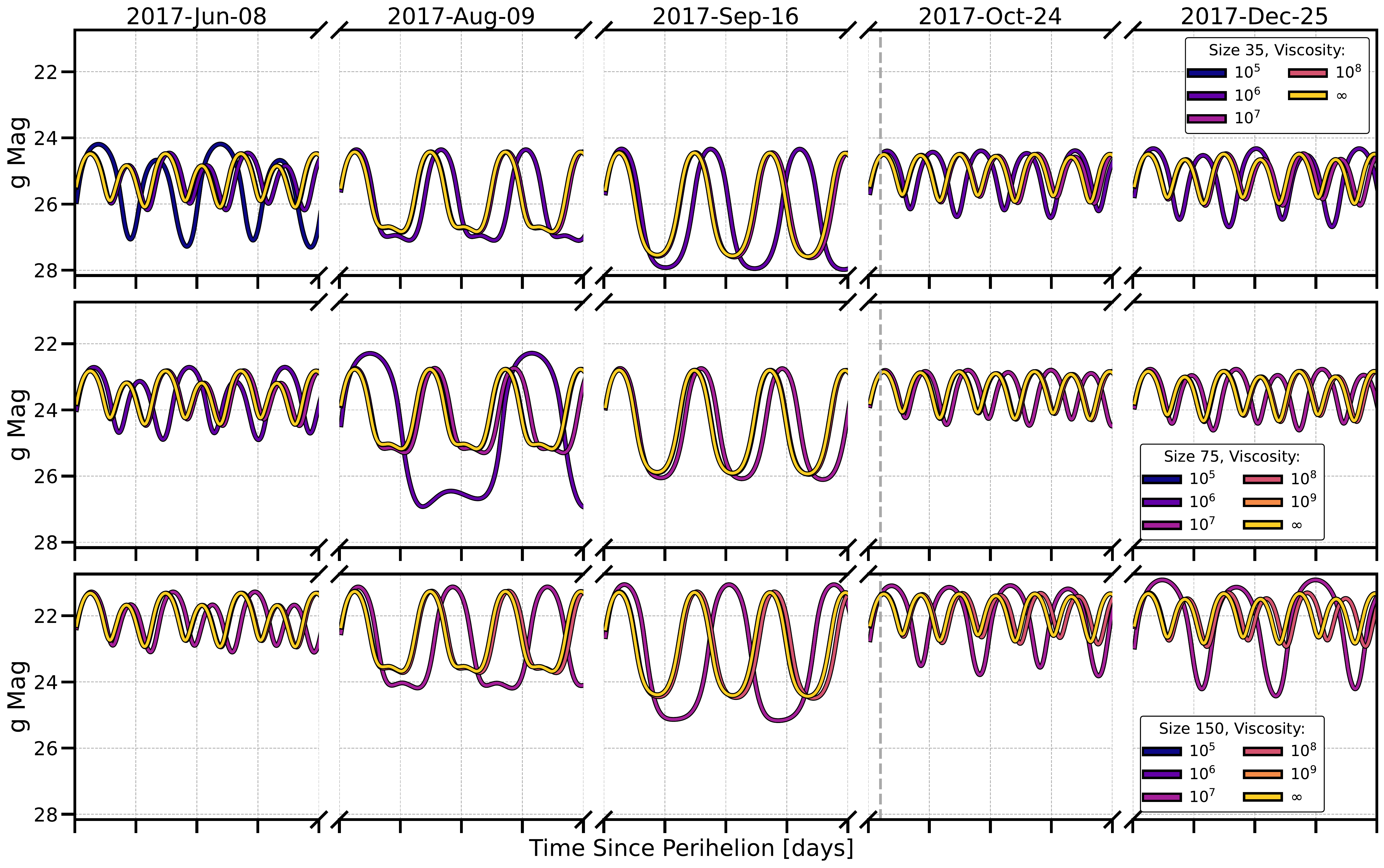}
\caption{Synthetic light curves of `Oumuamua which incorporate simulated changes in period and aspect ratio from tidal deformation. Rows correspond to the initial sizes and colors correspond to the dynamic viscosities of the body in the simulations. The time of discovery is indicated with a vertical dashed line.}
\label{fig:optimalaxislightcurvesims}
\end{figure*}

In Figure \ref{fig:optimalaxisheatmap}, we present an overview of all of the simulations. The moment of inertia is calculated as $I(t)=\int\rho \boldsymbol{r}(t)^2\text{d}x$, and we show $I(t_\text{max})/I(0)$. These changes in the moment of inertia will also affect the spin period. 

To quantify this effect, we assume an idealized scenario in which the spin period only changes in response to tidal deformation and subsequent changes in the moment of inertia. This is not realistic, because outgassing torques should dominate the rotational state. {However, we are only considering the tidal deformation of this object in the absence of outgassing.} We assume that the angular momentum is conserved such that $L(0)=L(t)$, where $L(t)=I(t) \omega(t)$ and $\omega(t)$ is the angular frequency. We additionally set $p(0)=7.3906$ hours, chosen such that the object has $p=7.3975$ hours at the time of detection, for the simulation which produced an optimal fit to the photometric data. The evolution of the spin period is shown in Figure \ref{fig:optimalaxisperiods} and the aspect ratios in Figure \ref{fig:optimalaxisaspectratio}. For high viscosities, the aspect ratios and the spin period are approximately constant, as the viscous forces are much stronger than the {tidal} forces. The rotational period increases in every case, since {the rotation about the z-axis forces the tidal deformation to} increase the moment of inertia. Although these simulations are highly idealized and do not incorporate outgassing {torques}, it is clear that tidal deformation can change the rotational state and light curve amplitude, based on the viscosity. 

\begin{table*}[t]
\begin{tabular}{ |c|c|c|c|c|c|c| } 
 \hline
 \multicolumn{7}{||c||}{$\chi^2$ of SIMULATED LIGHT CURVES vs DATA} \\
 \hline\hline
 \multicolumn{1}{|c|}{Size [m]} & \multicolumn{6}{c|}{Dynamic Viscosity [g cm$^{-1}$ s$^{-1}$]} \\
 \hline\hline
 \multicolumn{1}{|c|}{} & $10^5$ & $10^6$ & $10^7$ & $10^8$ & $10^9$ & $\infty$ \\
 \hline
 35 & - & 5.038e+04 & 3.882e+04 & 3.833e+04 & - & 3.831e+04 \\
 \hline
 55 & - & 1.078e+05 & 3.981e+04 & 3.840e+04 & - & 3.835e+04 \\
\hline
75 & - & - & 4.175e+04 & 3.851e+04 & 3.830e+04 & 3.829e+04 \\
\hline
95 & - & - & 4.539e+04 & 3.867e+04 & 3.832e+04 & 3.830e+04 \\
\hline
115 & - & - & 5.217e+04 & 3.886e+04 & 3.833e+04 & 3.831e+04 \\
\hline
150 & - & - & 8.213e+04 & 3.935e+04 & 3.837e+04 & 3.833e+04 \\
\hline
\end{tabular}
\caption{ $\chi^2$ values for each synthetic light curve fit to the photometric data. '-'s indicate simulations which were numerically unstable.}
\label{table:simdatachi2}
\end{table*}

Overall, objects with larger sizes undergo significantly more deformation and require larger viscosities to maintain stability. A dynamic viscosity of $\mu=10^7$ g cm$^{-1}$ s$^{-1}$ is sufficient for all {tested sizes to maintain physical conditions} over the trajectory. For initial sizes of 35 and 55 meters, $\mu=10^8$ g cm$^{-1}$ s$^{-1}$ lead to only small-scale changes in the body over its trajectory. For the remaining sizes (75, 95, 115, and 150 meters), $\mu=10^9$ g cm$^{-1}$ s$^{-1}$ similarly allows only minor changes. 

\subsection{{Synthetic Light Curves Incorporating Tidal Effects}}\label{subsec:lightcurvesim}

In this section, we present synthetic light curves for `Oumuamua over 5 1-day periods which incorporate the effects of tidal deformation. These 5 1-day periods begin on 2017 May 4, 2017 August 9, 2017 September 15 (perihelion), 2017 October 24 (detection), and 2018 January 29 (final simulated point). As before, trajectory and phase angle data are obtained from \href{https://ssd.jpl.nasa.gov/horizons.cgi}{JPL's Horizons database}. 

The light curves are computed with Equation \ref{eq:MLmodel} (the ML15 model) assuming optimal {``}fixed-axis{"} NPA rotation parameters (Section  \ref{subsec:gettingaxis}). The ML15 model incorporates a Lommel-Seeliger scattering surface for an ellipsoidal body, phase angle effects, and an arbitrary orientation and rotation. We compute the period evolution using the time evolution of the aspect ratio and moment of inertia (Figures \ref{fig:optimalaxisperiods} and \ref{fig:optimalaxisaspectratio}) computed in Section \ref{sec:oumuamuasim}. This period is used to compute the orientation, $\beta$, as $\beta=2\pi\,(t\%p)/p$, {with `\%' representing the modulo}. The ML15 model also incorporates the instantaneous aspect ratio into the light curve, {using simulated data provided by \texttt{SAMUS}}. The final synthetic light curves that incorporate amplitude and period modulations are presented in Figure \ref{fig:optimalaxislightcurvesims}. For comparison, we also show light curves with constant period and aspect ratio, which correspond to the dynamic viscosity $\mu\rightarrow\infty$.

Each model is initialized on 2017 May 4 with period $p=7.3975$ and aspect ratio 6:6:1. The parameters for the observation vector and the {``}fixed-axis{"} NPA rotation are the optimal results from Section \ref{subsec:oumuamuaaxis}. As these simulations use initial sizes and rotation periods which were obtained from data measured in October 2017, these light curves do not match up well with available photometric data, but instead provide qualitative examples of an evolving light curve due to tidal effects. 

These light curves are corrected for brightness variation due to {`Oumuamua}'s helio- and geo-centric distances. Therefore, amplitude variations are due to modulations of the aspect ratio and/or phase angle. Larger objects have brighter average magnitudes because the albedo is constant in all simulations. There are curious features in the light curves in August, which are due to changes in phase angle (Section  \ref{subsec:phaseaspectanalysis}). 

Aside from the effects of phase angle and period modulations, there is little variation in the shape of these light curves. Additionally, there is a notable decrease in amplitude post-perihelion. While the aspect ratio does increase during this time period, this feature in the light curve is entirely due to the phase angle. This effect is most obvious for the low-viscosity simulations. This implies that the evolving aspect ratio does not produce an observable signature in the amplitude for these simulation parameters. While tidal deformation can significantly alter the spin period, its effect on the amplitude of the light curve is not detectable for `Oumuamua. Period and amplitude changes in light curves of future interstellar objects may result from tidal deformation, and could be used to constrain the dynamic viscosity, and potentially the material composition. 

\begin{figure}[ht]
\centering
\includegraphics[width=.98\linewidth,angle=0]{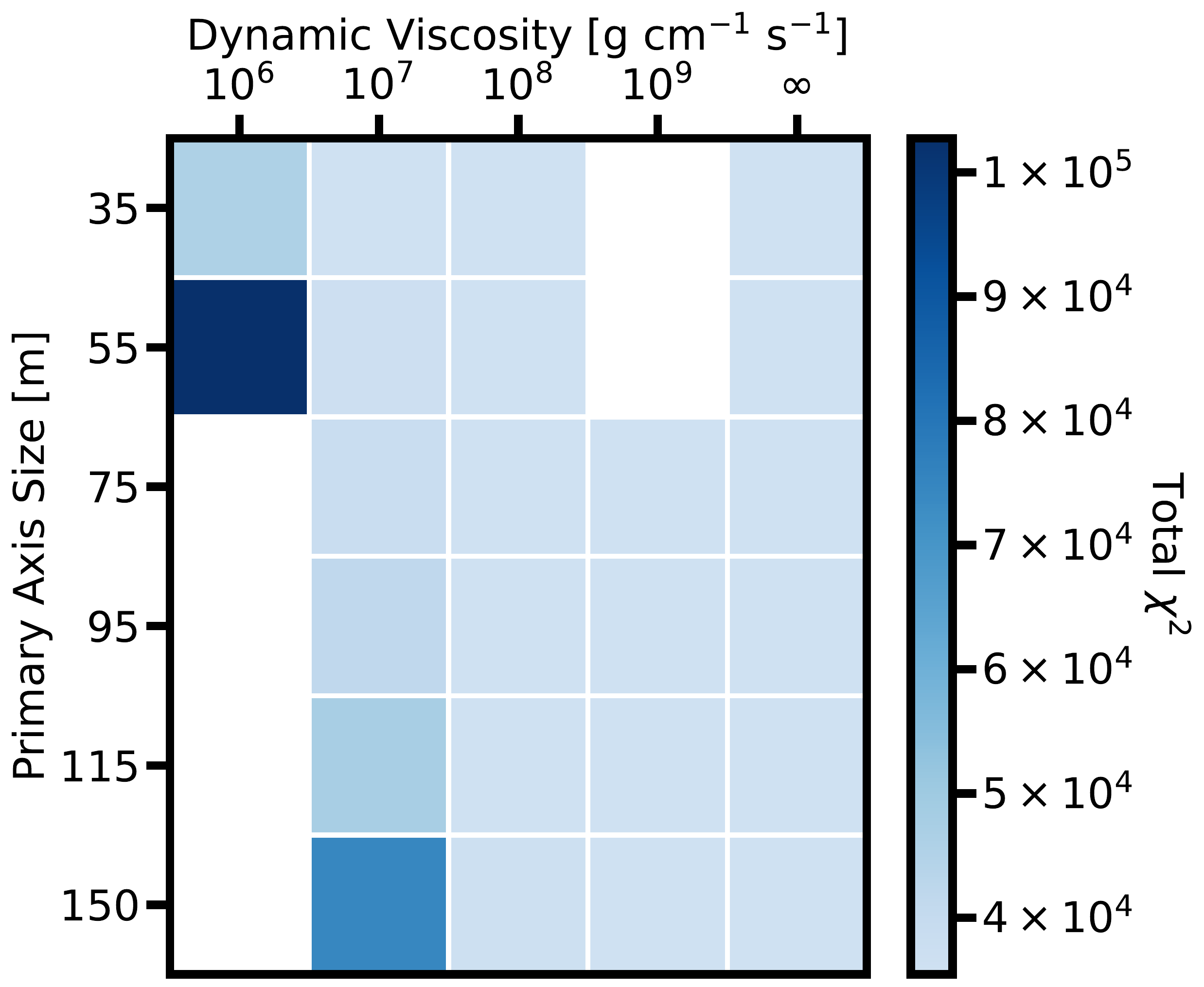}
\caption{Best fit $\chi^2$ values between synthetic and photometric data. Empty spaces indicate parameter choices for simulations that were not run. {Fits for both October and November are both computed and the $\chi^2$ values are added together.}}
\label{fig:simdatachi2heatmap}
\end{figure}

\begin{figure*}
\centering
\includegraphics[width=.94\textwidth,angle=0]{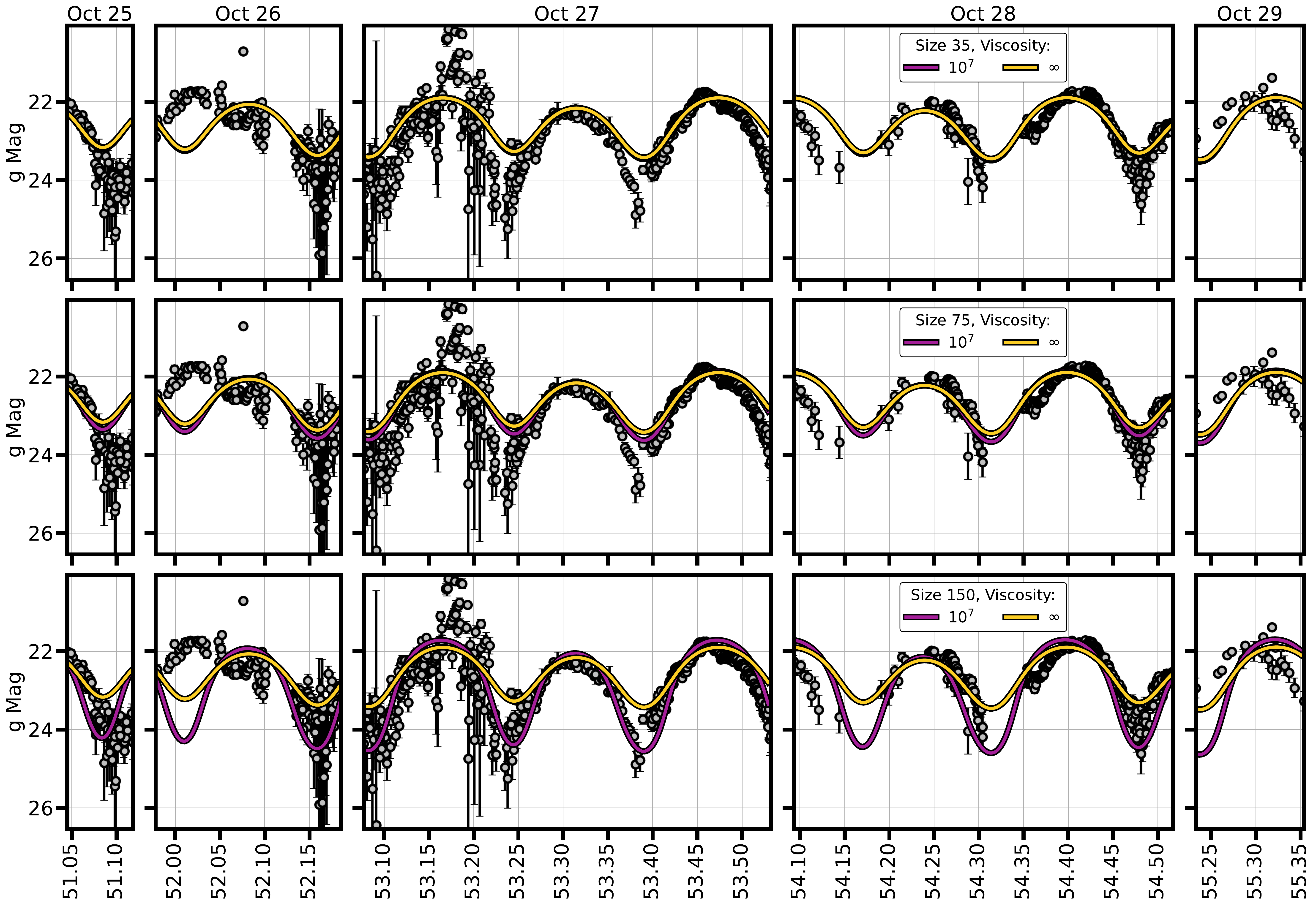}
\caption{Synthetic light curves for `Oumuamua which incorporate simulated aspect ratio changes and constant period. Colored lines indicate the dynamic viscosity and photometric data are shown in grey points. These curves use optimized values of $\beta_0$ and $\Delta V$. {Note that the rotation state is entirely fixed in these curves, with no evolution in the rotation period or axis. This allows for comparison to the photometric data and the original composite light curve, since the evolution of the rotation will require a more complex and physically accurate model. }} 
\label{fig:optimalaxiscurvesimdata}
\end{figure*}

We also compare the synthetic light curves with the photometric data, using the optimal period of $p=7.3975$ hours. As in Section  \ref{sec:lightcurvefits}, we optimize the initial rotation state $\beta_0$ and constant $\Delta V$ with \texttt{scipy.optimize.curve\_fit}, although we keep the remaining parameters constant. {By keeping the rotation state constant, we are able to compare these light curves to the optimal model computed in Section \ref{subsec:oumuamuaaxis}, despite the non-physical rotation model.} The November data is optimized independently, assuming $p_\text{Nov}=7.1910$, {which is found by a separate minimization}. In order to evaluate the validity of each fit, the $\chi^2$ values for both months are added together, which we present for each simulation in Table \ref{table:simdatachi2} and Figure \ref{fig:simdatachi2heatmap}. The resulting synthetic light curves along with the photometric data are shown in Figure \ref{fig:optimalaxiscurvesimdata} for October nights. 

\begin{figure*}
\centering
\includegraphics[width=\linewidth,angle=0]{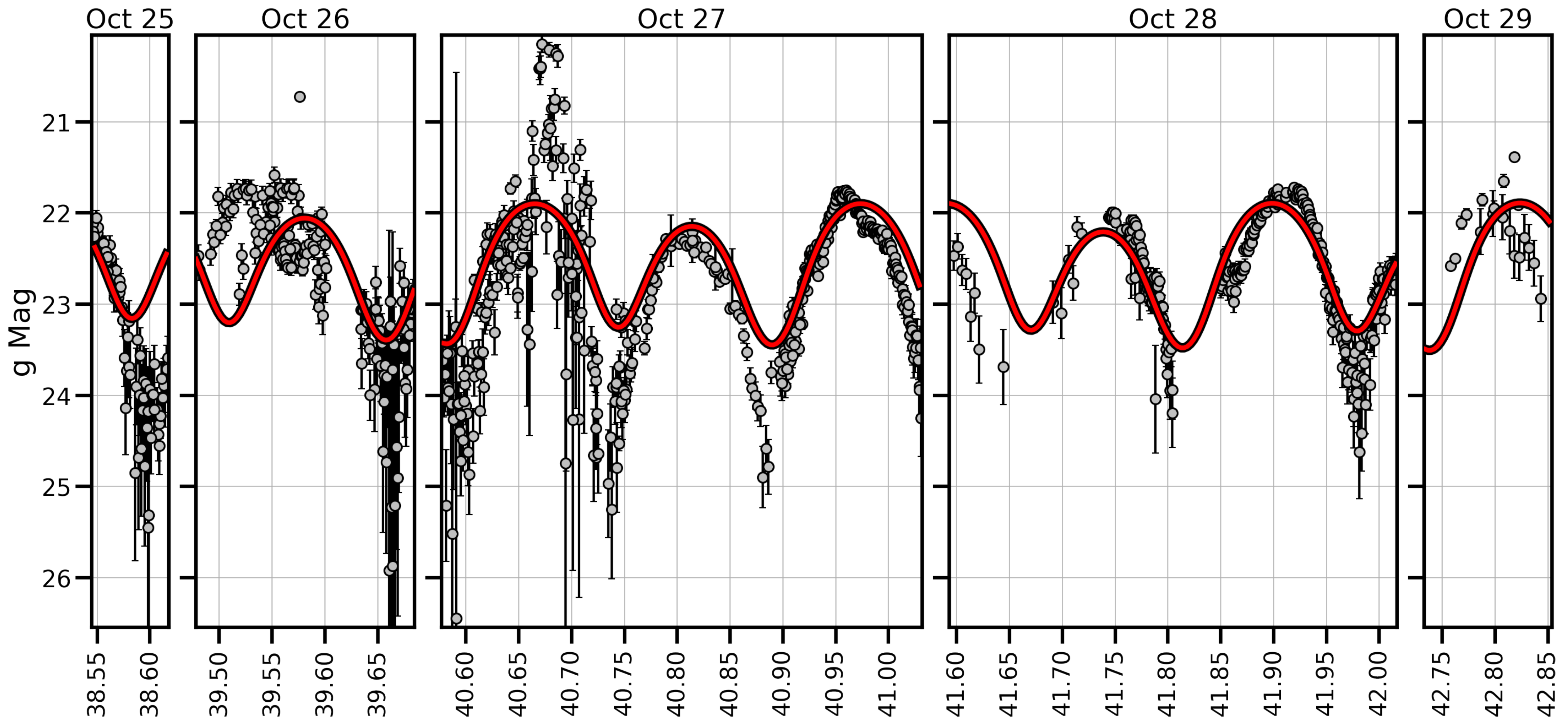}
\caption{The best-fit synthetic light curve (red line) and photometric data (grey points). This model incorporates the simulated changes in aspect ratio. Here $\mu=10^9$ g cm$^{-1}$ s$^{-1}$ and the body is initially 75 meters.}
\label{fig:optaxbestfitlightcurvesims}
\end{figure*}

A dynamic viscosity of $10^9$ g cm$^{-1}$ s$^{-1}$ and initial size of 75:75:12.5 meters produce optimal fits (discounting $\mu=\infty$), and this best-fit synthetic light curve is plotted along with the photometric data in Figure \ref{fig:optaxbestfitlightcurvesims}. It is worth noting that higher viscosity always produces a better fit for every initial size, therefore, no deformation provides the optimal explanation for `Oumuamua's light curve.
 
\section{Discussion}\label{sec:discussion}
 
In this paper, we {presented a novel open-source software (\texttt{SAMUS}) which models the deformation of ellipsoidal minor bodies under tidal stresses. This code is capable of constraining the material viscosity of such objects, as this deformation can produce observable changes in the rotation state and the photometric light curve. As a demonstration of the functionality of \texttt{SAMUS}, we applied it to the interstellar object 1I/`Oumuamua and investigated the material viscosity.} Our simulations show that {tidal} deformation would not cause measurable amplitude variations for `Oumuamua. It is possible that tidal deformation could be detected in other objects, depending on their bulk properties and albedo. {The non-detection of tidal deformation for `Oumuamua indicates that tidal forces were not a significant factor operating within its body, likely due to the rigidity/viscosity of this object. }

We assume in these simulations {and in \texttt{SAMUS} that the subject (`Oumuamua)} has a constant dynamic viscosity. However, {`Oumuamua} was tumbling and receiving solar irradiation across the surface. {\citet{Fitzsimmons2018}} modelled the radial thermal profile of `Oumuamua, assuming a thermal conductivity of $k=100$ erg s$^{-1}$ cm$^{-1}$ K$^{-1}$, a heat capacity of $C=5.5\cdot10^6$ erg g$^{-1}$, and a density of $\rho=1$ g cm$^{-3}$ (see Figure 4 {of that publication}). Those authors found that although the surface of the body reached a maximum of $T\simeq600$K at perihelion, subsurface layers at depths $\geq$30 cm never reached a temperature greater than $T\simeq150$K. Therefore, it is feasible that there was a strong temperature gradient immediately below the surface, which was not present in the cometary core. While the assumption of constant dynamic viscosity is less applicable for the surface half-meter, the majority of the body has a minimal temperature gradient and likely a nearly-constant dynamic viscosity. 

We {used \texttt{SAMUS} to constrain the effective dynamic} viscosity and size {of `Oumuamua}. For the optimal 115:111:19 meter size found by \citet{mashchenko2019}, the dynamic viscosity must be $\mu\geq10^9$ g cm$^{-1}$ s$^{-1}$, roughly equivalent to the viscosity of bitumen pitch ($\mu\simeq 3\cdot10^9$ g cm$^{-1}$ s$^{-1}$) \citep{edgeworth1984}.\footnote{This experiment measured pitch over temperatures ranging from 9$^\circ$C to 30$^\circ$C.} Allowable dynamic viscosities are provided for different initial sizes in Table \ref{table:allowedviscosities}. `Stability' here describes cases with $<1\%$ change in the moment of inertia over the trajectory. These are extremely high viscosities --- for comparison, terrestrial fluids such as water or olive oil have viscosities of $\mu\simeq10^{-2}$ and $\mu\simeq1$ g cm$^{-1}$ s$^{-1}$, respectively \citep{crc2022,fellows2022}. However, the allowable viscosity range is compatible with a variety of terrestrial materials, including water ice ($\mu\simeq 10^{13}$ g cm$^{-1}$ s$^{-1}$) \citep{fowler1997} and the terrestrial mantle ($\mu\simeq 10^{22}$ g cm$^{-1}$ s$^{-1}$), although the temperature dependence of dynamic viscosity makes these direct comparisons difficult.

{Unfortunately, these constraints on material composition are weakened by the difficulty of measuring the viscosity of proposed materials, especially at the relevant temperatures. While water ice has a known dynamic viscosity of $\mu\simeq10^{13}$ g cm$^{-1}$ s$^{-1}$ (over long-term high-latitude terrestrial temperatures) \citep{fowler1997}, the viscosities of exotic solids such as CO, H$_2$, and N$_2$ ice are far more difficult to measure. \citet{vilelladeschamps2017} analyzed the dynamics of N$_2$ ice glaciers in the Sputnik Planitia of Pluto, and found that the observed polygonal structure is consistent with viscosities of $10^{15}$--$10^{17}$ g cm$^{-1}$ s$^{-1}$, although these values are not well-constrained and are highly uncertain. Due to experimental barriers, however, no measurements are available for the viscosity of H$_2$ or CO ice. Additionally, the temperature dependence of viscosity renders the extension of such measurements to vacuum difficult, as measurements at such temperatures are not generally available. Finally, the structural composition of and material mixing within the body further complicate the viscosity measurements. Despite these practical difficulties, constraining the dynamic viscosity provides a new methodology to potentially test proposed structures and compositions for `Oumuamua, future interlopers of this type, and Solar System minor bodies.} 

\begin{table}
\begin{tabular}{ |c|c| } 
 \hline
 \multicolumn{2}{||c||}{MINIMAL STABLE VISCOSITY} \\
 \hline\hline
 Primary Axis [m] & Viscosity [g cm$^{-1}$ s$^{-1}$] \\
 \hline\hline
 35 & $10^8$ \\
 \hline
 55 & $10^8$ \\
 \hline 
 75 & $10^9$ \\
 \hline
 115 & $10^9$ \\
 \hline
 150 & $10^9$ \\
 \hline
\end{tabular}
\caption{Stable dynamic viscosities for initial sizes of `Oumuamua.}
\label{table:allowedviscosities}
\end{table}

The second-largest dynamic viscosity remains relatively stable in all cases, with small changes in moment of inertia slightly larger than the $<1\%$ simulation cutoff. These viscosities allow for changing period and amplitude, while preventing nonphysical divergence of the body. It should be noted, however, that the viscosity of $10^9$ g cm$^{-1}$ s$^{-1}$ is orders of magnitude lower than the $10^{13}$--g cm$^{-1}$ s$^{-1}$ viscosity of water ice. Therefore, this constraint provides little differentiation between various solid ices.

{It is also worth noting that changes in the moment of inertia can arise from sources other than tidal deformation, including ice crystallization, nucleus size changes, and dense mantle formation \citep{watanabe1992}. Like in tidal deformation, these changes can cause the rotation state to detectably evolve. However, many of the effects will cause a decrease in the moment of inertia, in contrast to the increase that tidal deformation generally causes.}

{We have also} demonstrated that for {these models,} an initial size of 75 meters and a dynamic viscosity of {$\mu\gg10^9$} g cm$^{-1}$ s$^{-1}$ {(no deformation)} provides an optimal match to the photometric data, {although different object sizes fit nearly equally well, and are relatively unconstrained by this model. These results indicate that tidal deformation likely played little-to-no role in the rotation state or the shape of `Oumuamua along its Solar System trajectory. It is possible that the change in period detected by \citet{flekkoy2019} is due to non-tidal torques.} {While this result is not strongly dependent on the rotation (see \href{https://github.com/astertaylor/Oumuamua}{GitHub} for simulations around the principal axes), a fully physical NPA rotation model would be needed to confirm this result.}

`Oumuamua left many unanswered questions as it exited the Solar System, and despite intense scrutiny, there is still no general consensus regarding the provenance of the object. The discovery implies a spatial number density of similar objects of order $n_{o}\sim1-2\times 10^{-1}\,$au$^{-3}$ \citep{Trilling2017,Laughlin2017,jewitt2017,moro2018,Zwart2018,Do2018,moro2019a}. Detection and characterization of future interstellar objects offer the most promising avenue for resolving these questions. 

The forthcoming Rubin Observatory Legacy Survey of Space and Time (LSST) \citep{jones2009lsst,Ivezic2019} will effectively detect such transient objects \citep{solontoi2011comet,Veres2017,veres2017b,Jones2018}. The survey should detect $\sim$1 `Oumuamua-like interstellar object every year \citep{Moro2009,Engelhardt2014,Cook2016,Trilling2017,Seligman2018,Hoover2022,Marceta2023}. In addition, the forthcoming NEO Surveyor \citep{Mainzer2015} may also detect interstellar objects, and could offer information about outgassing sources via its infrared capabilities. Space based in-situ measurements of an interstellar object would provide valuable information regarding the composition and bulk properties \citep{Hein2017,Seligman2018,Meech2019whitepaper,Castillo-Rogez2019,jones2019,Hibberd2020,Donitz2021,Sanchez2021,Meech2021,Hibberd2022,Moore2021whitepaper,Moore2021}. Additionally, future observations of the amplitude variations (if any) of interstellar objects may be used to constrain their viscosities --- and may be able to constrain composition --- via the techniques developed in this paper.

{In the future, similar analyses of tidal deformation with \texttt{SAMUS} may be useful for other small bodies. For objects with high rotation rates and closer solar approaches, tidal and centrifugal forces will have a larger effect. An example is 3200 Phaethon which also exhibits unexplained activity \citep{Jewitt2010,Jewitt2013,Li2013,Hui2017b} and is a target for the Japan Aerospace Exploration Agency (JAXA) DESTINY+ mission \textit{in situ} mission \citep{Arai2021}. For objects like 3200 Phaethon, tidal deformation may be detectable in photometric data, and would enable more strenuous constraints on the bulk physical properties of those objects. The recent detection of `dark comets' \citep{Chesley2016,Farnocchia2022, Seligman2022} and their still-unknown provenance provides another class of objects which may benefit from the application of \texttt{SAMUS} and other techniques developed in this paper. }

{The 2029 near-Earth flyby of 99942 Apophis is also an excellent candidate for the application of \texttt{SAMUS}. We demonstrate that the significant tidal forces experienced by Apophis may lead to deformation in its shape, which could be detectable either by photometric observations, radar observations, or the OSIRIS-APEX mission. Using \texttt{SAMUS}, the magnitude of such deformation would provide constraints on the internal material properties of Apophis.}

However, it is critical to note that the rotation model used in  \texttt{SAMUS} is non-physical {``}fixed-axis{"} NPA rotation. Because of the tumbling rotation states of `Oumuamua and Apophis, a fully physical model will be necessary to fully address the evolution of `Oumuamua and to apply \texttt{SAMUS} to Apophis in the coming years.

{The implementation of additional factors into \texttt{SAMUS} is also a worthwhile subject for future work. Particularly relevant is the addition of a shifting and strictly-conserved angular momentum axis, under the influence of external torques and tumbling. This addition will enable more accurate simulations of the deformation and rotation of such objects and thereby more accurate light curve fitting and analysis. The addition outgassing torques and ablation to \texttt{SAMUS}, {and a physical rotation model} will further enable effective analysis of rotation states and material properties of minor bodies. }

\begin{acknowledgements}
\section{Acknowledgements}
 We thank Faith Vilas for useful advice regarding the scientific content and structure of this manuscript. We thank the two anonymous reviewers for insightful and helpful suggestions which greatly strengthened the scientific content of this manuscript. This research utilized the University of Chicago’s Research Computing Center for numerical calculations. We thank Adina Feinstein for useful conversations and suggestions. We thank David Jewitt for useful feedback on the originally submitted version of the manuscript. DZS acknowledges financial support from the National Science Foundation Grant No. AST-17152, NASA Grant No. 80NSSC19K0444 and NASA Contract NNX17AL71A from the NASA Goddard Space Flight Center. 
\end{acknowledgements}

\bibliography{main}{}
\bibliographystyle{aasjournal}

\appendix

\section{{Light Curve Model Validation}}\label{sec:simplemodel}

\begin{figure}
\centering
\includegraphics[width=.5\linewidth,angle=0]{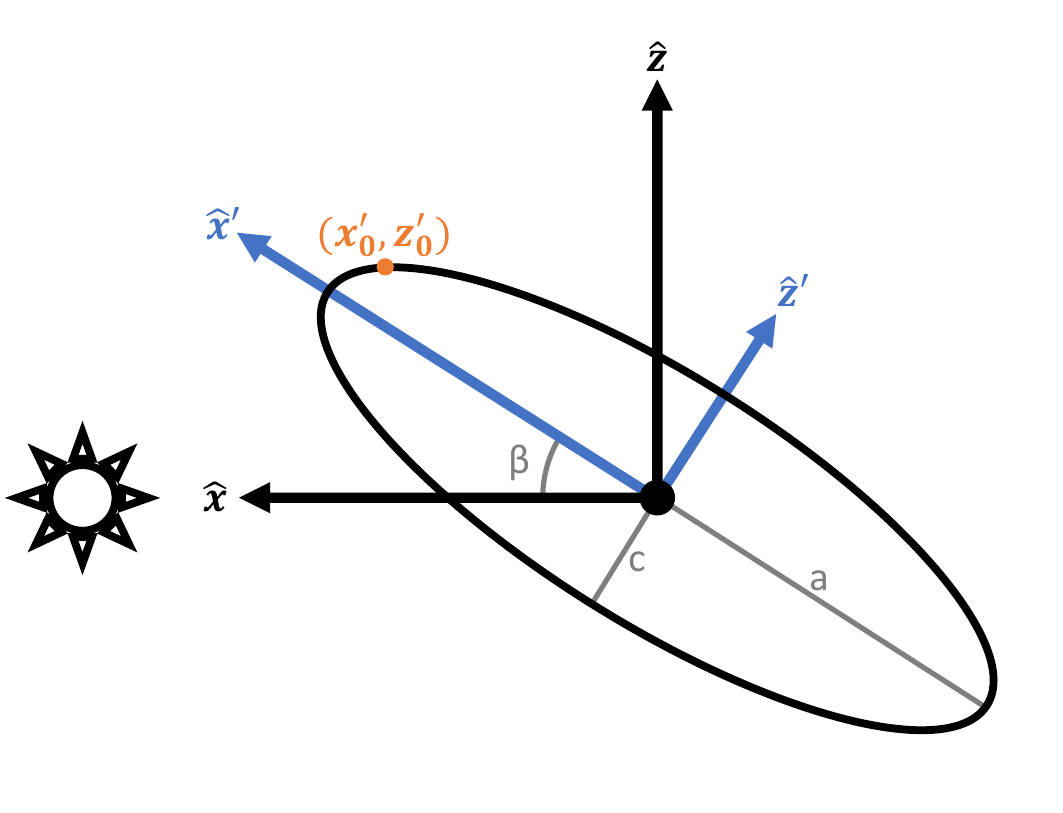}
\caption{{Diagram of the cross-sectional ellipse. Unprimed coordinates are fixed in the reference frame of the Solar System, while primed coordinates are fixed in the reference frame of the object. The angle $\beta$ defines the rotation state of the body with respect to the Sun. }}
\label{fig:ell-cross-section}
\end{figure}

{In this subsection, we validate the use of our ML15 model (described in Section \ref{sec:lightcurvefits}) via comparison to a simple principal-axis rotation that the complex model can be easily reduced to. To accomplish this, we produce a synthetic photometric time series of `Oumuamua assuming principal-axis rotation, a simplified viewing geometry, and no modulation from phase angle. We position the Sun, the Earth, and `Oumuamua along the x-axis in conjunction, roughly representative of the astrometric orbital arrangement when `Oumuamua was discovered. The non-rotating unprimed (x, y, z) axes move with `Oumuamua along its trajectory, while the primed coordinate axes (x', y', z') align with the principal axes and co-rotate with the body. We orient the body such that its semi-major principal axes are a:b:c$\sim$6:6:1 \citep{mashchenko2019}, and we assume that `Oumuamua rotates solely about the y-axis. Rotation about the x-axis produces a flat light curve, since the Sun-pointing projected area is unchanged. Due to `Oumuamua's (assumed) x-y symmetry, rotation about the z-axis similarly produces no change in the projected area and a flat light curve, leaving the y-axis as the only non-trivial principal axis rotation.} 

{We derive the projected area of an ellipsoid with an arbitrary rotation angle (with respect to the x-axis) $\beta$. It is known that any cross-section of an ellipsoid is an ellipse, although possibly with different aspect ratio and orientation. With rotation about the y-axis, the principal axes of the ellipse are along the Cartesian axes, and we solve for the lengths of these 2-dimensional principal axes. The semi-major axis in the y-direction is $b$, and so we now solve for the semi-major axis in the x-z plane which is orthogonal to the rotation axis. In this plane, the cross-section has an aspect ratio of 6:1, with $a=6$ in the x'-direction and $c=1$ in the z'-direction (Figure \ref{fig:ell-cross-section}). The observation vector, defined to be parallel to $\boldsymbol{\hat{x}}$, is rotated about the y-axis such that the rotated axis is $\cos\beta\,\boldsymbol{\hat{x}'}+\sin\beta\,\boldsymbol{\hat{z}'}$, in the primed frame. We then find the location of the points $(x'_0,z'_0)$ where the tangent to the ellipse is parallel to the observation vector, whose projection along the z-axis is this semi-major axis.}

{The cross-sectional ellipse is defined by the function $f_\text{2d}(x',z')$, where 
\begin{equation}\label{eq:ellipsesimple}
	f_\text{2d}(x',z')=\frac{x'^2}{a^2}+\frac{z'^2}{c^2}-1.
\end{equation}
The tangent $\partial z'/\partial x'$ is the derivative with respect to x of Equation \ref{eq:ellipsesimple}, given by
\begin{equation}
	\frac{\partial z'}{\partial x'}(x)=\mp \frac{c x'}{a^2\sqrt{1-x'^2/a^2}}\,.
\end{equation}
The $\mp$ reflects the degeneracy of the ellipse along slices in the x' axis. The slope of the observation vector in the x'-z' plane is simply $\tan(\beta)$, so by setting $\partial z'/\partial x'=\tan(\beta)$, and simplifying, we obtain 
\begin{equation}\label{eq:csfinalx}
    x'_0=\pm\frac{a^2\tan\beta}{\sqrt{c^2+a^2\tan^2(\beta)}}\,.
\end{equation}
Now by substituting Equation \ref{eq:csfinalx} into $z'=\pm c\sqrt{1-x'^2/a^2}$ (derived from Equation \ref{eq:ellipsesimple}), the points $(x'_0,z'_0)$ are given by
\begin{equation}\label{eq:x0z0}
    (x'_0,z'_0)=(\pm\frac{a^2\tan\beta}{\sqrt{c^2+a^2\tan^2(\beta)}},\mp \frac{c^2}{\sqrt{c^2+a^2\tan^2(\beta)}}).
\end{equation}}

{The distance between these points and the observation line which passes through the origin is the second semi-major axis of the projected ellipse. For a point $(x'_0,\ z'_0)$, the distance is the projection onto the z-axis, so $d=|\sin\beta\, x'_0+\cos\beta\, z'_0|$. Substituting Equation \ref{eq:x0z0}, the distance is
\begin{equation}
\begin{aligned}
	d=\frac{|a^2\sin\beta\tan\beta+c^2\cos\beta|}{\sqrt{c^2+a^2\tan^2(\beta)}}\,.
\end{aligned}
\end{equation}
Since the area of the ellipse is $\pi b d$, the brightness $L_\text{fix}$ (with simplification) is given by
\begin{equation}\label{eq:fixedlightcurve}
	L_\text{PA}=\pi b\sqrt{(a\sin\beta)^2+(c\cos\beta)^2}. 
\end{equation}
We verified (not shown) that this simple analytic model roughly reproduces the photometric data given in \citet{belton2018}.}

To compare and validate the ML15 model versus this one, we computed the error between the ML15 model and the principal-axis model described in this section. The error between these light curves is at machine precision. Similarly, we constructed an additional, numeric model allowing for {``}fixed-axis{"} NPA rotation, while maintaining the fixed and simplified astrometric arrangement. The error between this model and ML15 is of $\mathcal{O}(10^{-5})$. These comparisons validate the use of the ML15 model in this paper. 

\section{{Pseudocode}}\label{sec:pseudocode}
 
{In this subsection we present a pseudocode which demonstrates the most basic functionality of \texttt{SAMUS}. As a class method, \texttt{SAMUS} is capable of running in a modular format with complex function call patterns. In steps 1-8, we initialize the simulation. In steps 9.a-9.d, we compute the forcing functions described in Section \ref{subsec:SAMUStidalforce}, Table \ref{table:forcing}, and Equation \ref{eq:gaussgrav}. In steps 9.e-9.i, we update the model for a single time step. In steps 10 and 11, we implement the trajectory jump method described in Section \ref{subsec:trajjump}, and in step 12, we save and output simulation products. A script to replicate this example is available in the \texttt{SAMUS} package. }

\begin{enumerate}
\itemsep0em
    \item[]\texttt{SAMUS Pseudocode}
    \item\texttt{Set parameters. }
    \item\texttt{Read in 3-ball mesh and reshape to ellipsoid.}
    \item\texttt{Read in trajectory data.}
    \item\texttt{Create functions and function spaces for all variables.}
    \item\texttt{Create a UFL form for the Navier-Stokes equations.}
    \item\texttt{Create a FEniCS non-linear solver for the Navier-Stokes equations.}
    \item\texttt{Create a FEniCS non-linear solver for the Gaussian gravity formulation.}
    \item\texttt{Set t=0.}
    \item\begin{description}
            \item[WHILE]\verb|Number of cycles < number in loop:|
            \begin{enumerate}
                \item\texttt{Update tidal force, computed using Eqn }\ref{eq:samustide}.
                \item\texttt{Update gravitational force, computed using Eqn }\ref{eq:gaussgrav}.
                \item\texttt{Update Coriolis force, computed using the relevant term from Table }\ref{table:forcing}.
                \item\texttt{Update centrifugal force, computed using the relevant term from Table }\ref{table:forcing}.
                \item\texttt{Solve Navier-Stokes equations. If this diverges, STOP.}
		\item\texttt{Move mesh with velocity.}
                \item\texttt{Check if deformation crosses threshold. If it does, STOP.}
		\item\texttt{Compute moment of inertia using Equation \ref{eq:MOI}.}
		\item\texttt{Update time step.}
            \end{enumerate}
	\end{description}
    \item\texttt{Average velocities over the rotation cycles.}
    \item\begin{description}
            \item[WHILE]\texttt{Change in the heliocentric distance is less than tolerance:}
            \begin{enumerate}
                \item\texttt{Check to ensure that CFL<C}$_{\texttt{max}}$.
                \item\texttt{Move mesh using average velocity}.
                \item\texttt{Update time step.}
            \end{enumerate}
	\end{description}
    \item[\textbf{IF}:]\texttt{The number of steps is less than the upper limit, then repeat Steps 8-11.}
    \item\texttt{Save the mesh, the functions, and the moments of inertia over the path.}
\end{enumerate}

{This pseudocode structure is also shown in a flowchart available in the \texttt{SAMUS} documentation \citep{SAMUS}.}

\end{document}